1    Vesicularity, bubble formation and noble gas fractionation during MORB degassing.

2                     Geoffroy J. Aubry, Nicolas Sator and Bertrand Guillot

3      *Laboratoire de Physique Théorique de la Matière Condensée, Université Pierre et Marie Curie (Paris 6),*

4                  *UMR CNRS 7600, case courrier 121, 4 place jussieu,  75252 Paris cedex 05, France.*

8                                              **ABSTRACT**

9    The objective of this study is to use molecular dynamics simulation (MD) to evaluate the vesicularity

10   and noble gas fractionation, and to shed light on bubble formation during MORB degassing. A

11   previous simulation study (Guillot and Sator (2011) GCA 75, 1829-1857) has shown that the solubility

12   of $CO_2$ in basaltic melts increases steadily with the pressure and deviates significantly from Henry's

13   law at high pressures (e.g. ~9.5 wt% $CO_2$ at 50 kbar as compared with ~2.5 wt% from Henry's law).

14   From the $CO_2$ solubility curve and the equations of state of the two coexisting phases (silicate melt and

15   supercritical $CO_2$), deduced from the MD simulation, we have evaluated the evolution of the

16   vesicularity of a MORB melt at depth as function of its initial $CO_2$ contents. An excellent agreement is

17   obtained between calculations and data on MORB samples collected at oceanic ridges. Moreover, by

18   implementing the test particle method (Guillot and Sator (2012) GCA 80, 51-69), the solubility of

19   noble gases in the two coexisting phases (supercritical $CO_2$ and $CO_2$-saturated melt), the partitioning

20   and the fractionation of noble gases between melt and vesicles have been evaluated as function of the

21   pressure. We show that the melt/$CO_2$ partition coefficients of noble gases increase significantly with

22   the pressure whereas the large distribution of the $^4He/^{40}Ar*$ ratio reported in the literature is explained

23   if the magma experiences a suite of vesiculation and vesicle loss during ascent. By applying a pressure

24   drop to a volatile bearing melt, the MD simulation reveals the main steps of bubble formation and

25   noble gas transfer at the nanometric scale. A key result is that the transfer of noble gases is found to be

26   driven by $CO_2$ bubble nucleation, a finding which suggests that the diffusivity difference between *He*

27   and *Ar* in the degassing melt has virtually no effect on the $^4He/^{40}Ar*$ ratio measured in the vesicles.

34   **1. Introduction**

35   Noble gases are important tracers to understand the evolution and the degassing of the Earth's

36   mantle through time (Allègre et al., 1983; Harper and Jacobsen, 1996; Pepin, 2006). The elemental

37   fractionation of noble gases and their isotopes recorded in oceanic basalts (MORBs and OIBs), if

38   properly deciphered, may give information on the source region and on the details of noble gas

39   transport up to the surface. For instance the difference in $^3He/^4He$ ratios between MORBs and OIBs is

40   at the basis of the canonical model in isotope geochemistry (Porcelli and Wasserburg, 1995; Graham,

41   2002) describing the lower mantle as undegassed (high $^3He/^4He$ ratio) and the upper mantle as

42   degassed (low $^3He/^4He$ ratio). But this model is in conflict with other data (e.g. the He paradox,

43   Anderson, 1998) and is periodically revisited (Meibom et al., 2003; Parman et al., 2005; Albarède,

44   2008; Gonnermann and Mukhopadhyay, 2009). On the other hand, MORB and OIB glasses show a

45   large distribution of the $^4He/^{40}Ar*$ ratio (~1-1000, where $Ar*$ is corrected for air contamination) which

46   is interpreted as the signature of different degassing scenarios. Thus a low $^4He/^{40}Ar*$ ratio (~1) is

47   explained by a closed system degassing where a $CO_2$-saturated basaltic melt ($CO_2$ is the main volatile

48   component in oceanic basalts before $H_2O$, noble gases being present as traces) vesiculates by $CO_2$

49   exsolution and ascends with very little vesicle loss, up to the seafloor where volatiles are released (e.g.

50   Sarda and Graham, 1990). In contrast, a high $^4He/^{40}Ar*$ ratio can be accounted for either by a Rayleigh

51   distillation process (Burnard, 1999; Moreira and Sarda, 2000; Colin et al., 2011), by a suite of

52   vesiculation and vesicle loss during magma ascent (Sarda and Graham, 1990; Sarda and Moreira,

53   2002) or by a kinetic disequilibrium occurring just before eruption (Aubaud et al., 2004). However

54   $CO_2$ has a very low solubility in basaltic melts at pressure corresponding to the seafloor (~0.5 ppmw at

55   one bar, Dixon (1997)) and therefore an overwhelming majority of erupting lavas are strongly

56   degassed and have lost their pristine volatile contents.

57   Remarkable exceptions are *popping rocks* (Hekinian et al., 1973) characterized by high $CO_2$

58   contents (~1. wt% $CO_2$ in the sample *2πD43*, Sarda and Graham (1990), Javoy and Pineau (1991)) a

59   large vesicularity (>10 % in volume), and a $^4He/^{40}Ar*$ ratio (~1.5 in *2πD43*) compatible with the 60  expected *K/U* ratio of the mantle (Allègre et al., 1987). These tholeiitic basalts likely have experienced

61  a $CO_2$ exsolution in the oceanic mantle and for this reason are considered as probing the source region

62  (Sarda and Graham, 1990; Javoy and Pineau, 1991). Nonetheless, a pending question is to know if

63  magmas even richer in $CO_2$ do exist and may exsolve a $CO_2$-rich fluid at greater depth in the mantle.

64  In this context, it is noteworthy that evidences of explosive eruptions at ocean spreading centers are

65  well documented (Hekinian et al., 2000; Eissen et al., 2003; Pineau et al., 2004; Sohn et al., 2008;

66  Clague et al., 2009; Shaw et al., 2010; Helo et al., 2011). The presence of extended volcaniclastic

67  deposits at great water depth (~4,000 m) suggests that this explosive volcanism with magma disruption

68  could be driven by $CO_2$-rich melts (Sohn et al., 2008; Helo et al., 2011). More generally, there are

69  growing evidences of the existence of $CO_2$-rich magmas in the upper mantle. Thus, seismological and

70  magnetotelluric data suggest that incipient melting may begin at depths of 150-300 km in the upper

71  mantle (Evans et al., 2005; Bologna et al., 2011). Experimental petrology data shows that deep melting

72  can occur with carbonate-silicate assemblages leading to carbonatitic liquids at very low melt fraction

73  (Eggler, 1976; Dalton and Pressnall, 1998; Dasgupta and Hirschmann, 2006; Zeng et al., 2010;

74  Rooney et al., 2012). So, the electrical anomalies of the oceanic asthenosphere could be explained by

75  ~0.1 % volume fraction of carbonatite melts circulating in the silicate mantle (Gaillard et al., 2008).

76  This melt fraction corresponds on average to ~300 ppmw $CO_2$ stored in the asthenosphere, a value

77  compatible with various estimates of the $CO_2$ contents in the source region of MORB (Marty and

78  Tolstikhin, 1998; Saal et al., 2002; Cartigny et al., 2008). Furthermore these carbonatites can

79  metasomatize the surrounding silicate matrix and lead to melt mixing during their ascent (Dixon et al.,

80  2008).

81      When an upwelling volatile-bearing melt reaches the $CO_2$ saturation limit, the excess of $CO_2$ is

82  exsolved by forming bubbles or vesicles. The noble gases initially present in the melt as traces (less

83  than a ppmw, Moreira et al., 1998) fractionate between melt and $CO_2$ bubbles in a proportion which

84  depends both on the vesicularity and on their respective solubility (or partition coefficient) in the two

85  coexisting phases (Jambon et al., 1986). When a tholeiitic basalt vesiculates at shallow depth under the

86  seafloor (e.g. a few kilometers), a simple way to estimate the noble gas partitioning is to assume that the fluid filling the vesicles is an ideal gas and that the solubility of noble gases in the basaltic melt is that measured at laboratory at low pressures (Henry's law). Because noble gas solubilities in basaltic melts are very low ($S_{He-Xe}$ ~$10^{-5}$ - $10^{-8}$ bar$^{-1}$, where $S$ is the inverse of Henry's constant, Jambon et al. (1986)), noble gases partition preferentially into the $CO_2$ bubbles, the heavier the noble gas the higher the fraction. This is the common view of closed-system noble gas degassing.

A different situation is encountered if the $CO_2$ saturation limit is reached at great depth in the mantle (e.g. in the oceanic astenosphere). The vesicles are filled with a very dense supercritical $CO_2$ fluid which limits the transfer of noble gases from the melt. A theoretical approach of this mechanism has been proposed by Sarda and Guillot (2005) and Guillot and Sarda (2006) by using a hard sphere model to describe the silicate melt, the fluid $CO_2$, and the incorporation of noble gases in the two coexisting phases. An important conclusion of this study is that the noble gas fractionation between melt and vesicles at high pressures deviates significantly from a simple evaluation based on noble gas solubility data at low pressures (Henry's law) and for which the deviation from ideality of the $CO_2$ fluid is neglected. Although this model is useful for theoretical guidance, there is a need for improvement. For instance, the solubility of $CO_2$ in silicate melts and especially in basalts is well documented at low pressures where Henry's law holds (Stolper and Holloway, 1988; Dixon et al., 1995; Jendrzejewsky et al., 1997; Botcharnikov et al., 2005) but is poorly known at high pressures above ~20 kbar (Pan et al., 1991; Brooker et al., 2001). Recently Guillot and Sator (2011) have evaluated by molecular dynamics simulation (MD) the pressure evolution of the $CO_2$ solubility in silicate liquids of various composition. In basaltic melts the calculated $CO_2$ solubility increases steadily above 20 kbar (where $X_{CO_2}$~1.6 wt%) and reaches ~20 wt% at 80 kbar, a value much higher than the one estimated from Henry's law ($X_{CO_2}^H$~0.5 ppmw per bar, see Dixon et al. (1995) ). These simulation results are in agreement with a few solubility data in melts of basaltic composition obtained from high-pressure partial melting experiments of silicate-carbonate assemblages (Hammouda, 2003; Thomsen and Schmidt, 2008).

Our objective is to evaluate by atomistic simulation, the pressure dependence of noble gas partitioning between a $CO_2$-saturated MORB melt and a coexisting $CO_2$ phase (closed system degassing), as also as the noble gas fractionation in melt and in vesicles during MORB degassing. For that we have implemented the *test particle method* (TPM) to calculate the chemical potential of a trace element (e.g. a noble gas) into a melt or a high density fluid ($CO_2$) modeled by MD simulation. This method has been recently used to evaluate the solubility of noble gases in high-pressure silicate liquids (Guillot and Sator, 2012). The theoretical framework and the simulation details are explained in section 2, and the results are presented and discussed in section 3. In section 4, the kinetics of bubble formation and noble gas transfer is investigated by devising a numerical experiment where a pressure drop is applied to a volatile bearing MORB melt. The implications of such a decompression experiment are discussed.

## 2. Method of calculation

### 2.1 Noble gas partitioning in a vesiculated MORB melt.

When a volatile bearing MORB melt is ascending under an oceanic ridge, it starts to degas when the ambient pressure becomes lower than the saturation pressure of the major volatile component present in the melt (Sparks, 1978; Bottinga and Javoy, 1990). In a MORB melt where $CO_2$ is the major volatile component and noble gases are traces, the latter ones redistribute between $CO_2$ bubbles and bulk melt. At given temperature (T) and pressure (P) conditions, the equality of the chemical potentials of a noble gas $i$ in the vesiculated melt and in the $CO_2$ bubbles leads to the following relationship,

$$\rho_m^i/\rho_v^i = e^{-(\mu_m^{i,ex}-\mu_v^{i,ex})/k_BT} = \gamma_m^i/\gamma_v^i \qquad (1)$$

where $k_B$ is the Boltzmann constant, $T$ the temperature, $\rho_m^i$ and $\rho_v^i$ are the number densities (number of atoms per unit volume) of the noble gas $i$ in the melt and in the vesicles ($CO_2$ bubbles), respectively, and where $\mu_m^{i,ex}$ and $\mu_v^{i,ex}$ are the excess chemical potentials of the noble gas in the corresponding phases (notice that the ideal parts of the chemical potentials cancel out because they are identical in the two phases). The quantity $\gamma_{m,v}^i = e^{-\mu_{m,v}^{i,ex}/k_BT}$ is named the solubility parameter in the corresponding 137     phase ($m$ or $v$). By introducing $V_m$ the volume of melt and $V_v$ the total volume of vesicles the Eq.(1)

138     leads to the following equality,

139     $$\frac{N_m^i}{N_v^i} = \left(\frac{V_m}{V_v}\right) \cdot \left(\frac{\gamma_m^i}{\gamma_v^i}\right) \qquad\qquad (2)$$

140     where $N_m^i$ is the number of noble gas atom in the melt and $N_v^i$ the one in the vesicles.

141     For a closed system degassing, $N_m^i$ and $N_v^i$ are related by the following mass conservation law,

142     $$N_0^i = N_m^i + N_v^i \qquad\qquad (3)$$

143     where $N_0^i$ is the number of noble gas atom $i$ in the $CO_2$-bearing MORB melt before vesiculation.

144     Following the derivation of Guillot and Sarda (2006), the final step consists to introduce Eq.(3) into

145     Eq.(2) and to make use of the definition of the vesicularity, $V^* = V_v/(V_v + V_m)$. After some elementary

146     algebra the numbers of noble gas atom $i$ in the melt and in the vesicles are equal to,

147     $$N_m^i = N_0^i \cdot \left(\frac{\gamma_m^i}{\gamma_v^i}\right) \cdot (1 - V^*) / \left[V^* + \left(\frac{\gamma_m^i}{\gamma_v^i}\right) \cdot (1 - V^*)\right] \qquad\qquad (4)$$

148     $$N_v^i = N_0^i \cdot V^* / \left[V^* + \left(\frac{\gamma_m^i}{\gamma_v^i}\right) \cdot (1 - V^*)\right]. \qquad\qquad (5)$$

149     The above expressions are exact at equilibrium (for any thermodynamic condition) and they replace

150     the ones obtained by Jambon et al. (1986) which are valid only at very low pressure when the $CO_2$

151     fluid filling the vesicles can be considered as ideal (in this case $\gamma_v^i = 1$) and $\gamma_m^i$ can be approximated

152     by the solubility constant $S^i$ in using Henry's law,

153     $$\gamma_m^i = \rho_m k_B T S^i \qquad\qquad (6)$$

154     where $\rho_m$ is the numerical density of the melt. However, the simplicity of expressions (4) and (5) is

155     somewhat misleading because to evaluate $N_m^i$ and $N_v^i$ at given (T,P) conditions, the vesicularity of the

156     degassing melt must be known and the solubility parameters $\gamma_m^i$ and $\gamma_v^i$ evaluated in the coexisting

157     phases at these very thermodynamic conditions.

158    To estimate the evolution of the vesicularity with the thermodynamic conditions one needs to know

159    the pressure (and temperature) behaviour of the $CO_2$ solubility in the MORB melt. As a matter of fact,

160    in using the equality of the chemical potentials of $CO_2$ in the melt and in the vesicles as also as the

161    mass conservation law for the $CO_2$ molecules, it can be shown that the vesicularity, $V^*$, fulfils the

162    following relationship,

163
$$\frac{V^*}{1-V^*} = \left(\frac{n_m}{n_v}\right) . \left[\frac{\frac{(1-W)}{W}}{\frac{(1-W_0)}{W_0}} - 1\right] \tag{7}$$

164    where $n_m$ and $n_v$ are the densities of the $CO_2$-saturated silicate melt and of the $CO_2$ fluid filling the

165    vesicles at (T,P) of interest, $W$ is the solubility of $CO_2$ (in $g_{CO_2}/g_{melt}$) in the silicate at (T,P)

166    conditions during magma ascent, and $W_0$ is the $CO_2$ content in the melt before degassing ($W < W_0$). If

167    exsolution only occurs when a degree of super saturation in $CO_2$ is reached (i.e. when magma ascent is

168    so rapid that chemical equilibrium cannot be achieved) then the term "-1" on the right hand side of

169    Eq.(7) may be replaced by "-α " where the super saturation ratio, $\alpha = X_w^{CO_2}(P,T)/W(P,T)$ , expresses

170    the ratio of the weight fraction of $CO_2$ molecules actually in the melt to the one expected at saturation.

171    In the following α (≥1) will be used as an adjustable parameter.

172    **2.2 Calculation of noble gas solubility parameters.**

173    As expressed by Eqs.(4) and (5), in a closed system degassing, the partitioning of a noble gas $i$

174    between melt and vesicles depends on the vesicularity, $V^*$, and on the ratio of the solubility

175    parameters, $\gamma_m^i/\gamma_v^i$ . In a seminal work, Widom (1963) has shown that the solubility parameter, $\gamma_s^i$, of a

176    solute $i$ at infinite dilution in a solvent $s$ can be written under the following form,

177
$$\gamma_s^i = e^{-\mu_s^{i,ex}/k_BT} = < e^{-\psi/k_BT} >_{N_s} \tag{8}$$

178    where $\psi = (U_{N_s+1} - U_{N_s})$ is the potential energy difference between a mixture composed of $N_s$

179    solvent molecules plus the solute particle (e.g. a noble gas atom) and the pure solvent. In the case

180    where the potential energy is pairwise additive (i.e. $U_{N_s+1} = \sum_{i<j=1,N_s+1} u_{ij}$ where i and j are two

181      particles and $u_{ij}$ the pair interaction energy), $\psi$ is nothing but the solute-solvent interaction energy

182      ($\psi = \sum_{j \in N_s} u_{1j}$ where $u_{1j}$ is the interaction energy between the solute particle 1 and the solvent

183      particle $j$). A key feature of Eq.(8) is that the canonical average $< e^{-\psi/k_BT} >_{N_s}$ is taken over the

184      configurations of the pure solvent, the solute particle acting as a ghost (or test) particle. In practice, the

185      solute (i.e. the noble gas atom) is inserted at random in the solvent configurations generated by MD

186      simulation and the Boltzmann factor, $< e^{-\psi/k_BT} >_{N_s}$, is evaluated by averaging over all insertion

187      events. However, when the solvent density is high, the noble gas atom randomly inserted has a high

188      probability to overlap a solvent molecule. The consequence is a strongly repulsive solute-solvent

189      interaction energy ($\frac{\psi}{k_BT} \gg 1$) leading to a vanishingly small contribution of this event to the average

190      ($e^{-\psi/k_BT} \ll 1$). What is needed for practical use of this method is a numerical recipe for detecting

191      quickly undesirable positions of the inserted noble gas atom and for locating cavities that can

192      accommodate it, cavities that appear and disappear at the mercy of the solvent density fluctuations. We

193      have implemented the sampling method of Deitrick et al. (1989) which makes the evaluation of the

194      solubility parameter by the test particle method (TPM) very effective. For a detailed description of the

195      method and a discussion of the statistical uncertainties, the reader is referred to the paper by Guillot

196      and Sator (2012) where the solubility of noble gases in silicate melts was evaluated in this framework.

197      To evaluate the noble gas solubility parameters $\gamma_m^i$ and $\gamma_v^i$ by the TPM one needs to generate by MD

198      simulation a suite of atomic configurations which are representative of the $CO_2$-bearing basaltic melt

199      and of the $CO_2$ fluid assumed to coexist with each other at a given state point (T,P). The accuracy of

200      the calculation relies on the force field used to describe the interactions between the atoms of the

201      phases under consideration. Six different kinds of interaction are involved: $CO_2$-$CO_2$ interaction,

202      silicate-silicate interaction, $CO_2$-silicate interaction, noble gas-$CO_2$ interaction, noble gas-silicate

203      interaction and noble gas-noble gas interaction. Notice that noble gases are trace elements in the

204      mantle (on average much less than a ppmw in MORBs) and therefore can be considered at infinite

205      dilution both in the MORB melt and in the $CO_2$ bubbles. For silicate-silicate, $CO_2$-$CO_2$ and $CO_2$-

206      silicate interactions we have used the pair potentials developed by Guillot and Sator (2007a,b; 2011) in their MD studies of dry and $CO_2$-bearing silicate melts of various composition (the corresponding potential parameters are listed in Table 1 of Guillot and Sator (2011) and are not reproduced here).

When $CO_2$ is incorporated into the basaltic melt, a reactive force field accounts for the chemical reactivity between the oxygens of the silicate and $CO_2$ molecules,

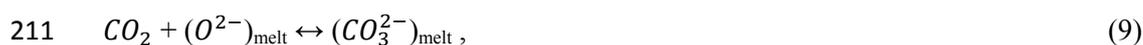

$$CO_2 + (O^{2-})_{melt} \leftrightarrow (CO_3^{2-})_{melt} , \qquad (9)$$

the equilibrium concentrations in $CO_2$ and carbonate ions ($CO_3^{2-}$) depending at once on the melt composition, the investigated thermodynamic conditions (T,P) and the total $CO_2$ content. In the case of a MORB composition, the $CO_2$ solubility curve has been investigated by Guillot and Sator (2011) at isothermal conditions (between 1673 and 2273K) up to 150 kbar. We have used these simulation data as benchmarks for preparing our $CO_2$+MORB compositions.

To evaluate the interaction energy between noble gases and the elements of the silicate melt, we

have implemented the atom-atom Lennard-Jones potentials recently developed by Guillot and Sator

(2012), a force field which successfully reproduces the solubility data of noble gases in liquid silicates.

As for the interactions between $CO_2$ and the noble gases we have developed specifically for the present study a force field based on the Lennard-Jones pair potential. The derivation of the potential parameters is presented in Appendix and their values are listed in Table A1.

**2.3 Simulation details**

Two series of MD runs were performed, one corresponding to a $CO_2$-saturated basaltic melt at various thermodynamic conditions, and the other one corresponding to the $CO_2$ phase assumed to be in coexistence with the melt. The composition of the simulated MORB (in wt%: 50.59 $SiO_2$, 1.5 $TiO_2$,

15.11 $Al_2O_3$, 1.15 $Fe_2O_3$, 8.39 $FeO$, 7.77 $MgO$, 11.87 $CaO$, 2.94 $Na_2O$ and 0.13 $K_2O$) is the one of a

sample (*TK21B*) of the mid-atlantic ridge. This composition has been investigated by Guillot and Sator

(2011) to determine the $CO_2$ solubility curve as function of T and P. So the $CO_2$ contents at saturation used in the present study are those determined by these authors at T=1873K and P=10, 20, 30, 50, 80,

and 100 kbar. With regard to the $CO_2$-saturated MORB melt, the simulation cell is composed of 1,000

232    ions (or 5,000 ions, to estimate system size effects) plus the necessary number of $CO_2$ molecules to

233    reach $CO_2$ saturation at the investigated state point (the simulation conditions are summarized in Table

234    1). The coexisting $CO_2$ phase is simulated independently from the $CO_2$-saturated melt and is composed

235    of 500 (or 2,500) molecules.

236    The MD calculations were performed with the DL_Poly 2.0 code (Smith and Forrester, 1996). The

237    equations of motion for ions and $CO_2$ molecules were solved with a time step of $10^{-15}$s. The simulation

238    box is cubic with periodic boundary conditions and the long range coulombic interactions are

239    accounted for by an Ewald sum. For the two systems (the $CO_2$ bearing melt and the $CO_2$ phase) the

240    calculations were first performed in the isothermal isobaric ensemble (N,P,T) for equilibration and

241    next were carried on in the microcanonical ensemble (N,V,E) for generating production runs. The

242    latter ones were $10^{-8}$s long (i.e. $10^7$ MD steps) for small system sizes (N=1,000 ions for MORB and

243    500 molecules for fluid $CO_2$), and $10^{-9}$s long for large system sizes (5,000 ions and 2500 molecules,

244    respectively). The statistical uncertainties on each investigated state point are about 1% for density,

245    pressure and temperature. At each state point under consideration, atomic configurations were stored

246    every $2.10^{-12}$s (i.e. 2,000 MD steps) to be sampled afterwards by the TPM (this time interval is chosen

247    so as the displacement of atoms between two successive recording is of the order of an atomic

248    diameter, i.e. ~3A).

249    As mentioned in section 2.2, the TPM was optimized in implementing the sampling procedure of

250    Deitrick et al. (1989). A description of this recipe is discussed in details in Guillot and Sator (2011)

251    and is not reproduced here. In short, the simulation box containing the atomic configurations to be

252    sampled is divided into small cubelets (e.g. 100×100×100). A cubelet is marked as occupied if its

253    center is located within the repulsive core of a solvent atom. Thus, only insertions where the test

254    particle is addressed to an unoccupied cubelet contribute to the average $< e^{-\psi/k_B T} >_{N_s}$. However, the

255    free volume accessible to the test particle (the volume fraction corresponding to unoccupied cubelets)

256    decreases drastically with the size of the test particle (the noble gas) to insert and with the density of

257    the solvent (the higher the density the smaller the free volume). For instance, the free volume accessible to He in a MORB melt at 10 kbar is 19%, whereas it is equal to 12% for Ne, 1.3% for Ar and only 0.04% for Xe. At 50 kbar the free volume amounts to 11% for He, 5.5% for Ne, 0.2% for Ar and 0.002% for Xe. Thus, the occurrence to find at the mercy of density fluctuations, a cavity large enough to accommodate an Ar atom in a basaltic melt at 50 kbar is a rare event. This is why it is necessary to perform long MD runs with a system size large enough. Moreover, it is important that the number of cubelets mapping the simulation box is large enough to be certain that all cavities susceptible to accommodate the noble gas atom have been sampled.

With regard to the MD simulations, the parameters to be fixed are the system size and the length of the MD run. When these two parameters are fixed, the numerical uncertainties associated with the evaluation of the solubility parameters ($\gamma_{m,v}^i$) by the TPM increase drastically with the size of the noble gas to insert and with the density (i.e. with the pressure) of the solvent (for a discussion see Appendix A in Guillot and Sator, 2012). To estimate these uncertainties we have performed two independent series of MD calculations which approximately lead to the same computational cost. One series was performed with small system sizes (see Table 1) over a long simulation time (10 ns) and another series with larger system sizes but over a shorter simulation time (1ns). The results of these calculations are presented in Table 2. A fair agreement is found between the two series of calculations for He and Ne in the two phases (the deviation is less than 3 percent) and for Ar in $CO_2$ (except at 100 kbar where $\Delta\gamma_v^{Ar}/\gamma_v^{Ar}\sim14\%$). However the deviation becomes more significant for Ar in the MORB melt above 50 kbar (e.g. $\Delta\gamma_m^{Ar}/\gamma_m^{Ar}\sim20\%$ at 50 kbar and $\sim40\%$ at 100 kbar) and is quite large for Xe in MORB from 30 kbar ($\sim300\%$ at 30 kbar). A drastic reduction of the uncertainties for Xe in MORB at high pressures would require to perform very long MD runs ($\sim1000$ ns) with a large number of atoms ($10^5$-$10^6$), a task which is too demanding for our numerical resources. Consequently values of $\gamma_m^{Xe}$ for pressure above 30 kbar will be discarded in the following. In considering that the calculations performed with the two system sizes are independent from each other, the final values of $\gamma_m^i$ and $\gamma_v^i$ have been obtained simply by doing the arithmetic mean of the two series of calculations.

### 3. Results

### 3.1 Vesicularity of MORBs

Before discussing the fractionation of noble gases between melt and vesicles it is worthwhile to analyze the pressure evolution of the vesicularity of a degassing magma (closed system degassing). In Fig.1 is reported the pressure dependence of the vesicularity of an ascending magma, described by our simulated MORB melt, at isothermal condition (1873 K) and for different initial conditions of $CO_2$ saturation at depth (with $P_{CO_2}^{sat}$ = 100, 80, 50, 30, 10, 5, and 1 kbar). The vesicularity, $V^*$, is evaluated from Eq.(7) in using for the pressure evolution of the $CO_2$ solubility ($W$) the one calculated by Guillot and Sator (2011), whereas the density ($n_m$) of the $CO_2$ bearing MORB melt and the density ($n_v$) of the coexisting $CO_2$ fluid are deduced from the present simulation study. By inspecting Fig.1, it appears that the growth rate of the vesicularity of an ascending magma is correlated with the $CO_2$ content of the source region (see in Fig.1 the evolution of the curvature of the vesicularity curve with increasing $P_{CO_2}^{sat}$). For instance, a MORB melt $CO_2$-saturated at $P_{CO_2}^{sat}$ = 10 kbar (d~30 km) will show a vesicularity of 10% in erupting at 5,500 m below sea level (m.b.s.l.) whereas the one of an ascending magma $CO_2$-saturated at $P_{CO_2}^{sat}$ = 100 kbar (i.e. d~300 km) will reach 10% at $P$ = 90 kbar (d~270 km) and 50% at 15 kbar (d~45 km). So, if it exists a source region very enriched in $CO_2$ in the oceanic mantle (from a carbonatitic origin, for instance), the magma issuing from this setting will become highly vesiculated at great depth. In that case the supercritical $CO_2$ fluid is characterized by a liquid-like density (e.g. $n_{CO_2}$~1.40 g/cm³ at 30 kbar) and a degassing process at this depth is more related to a liquid-liquid immiscibility than to a gas bubble formation.

Actually, an overwhelming majority of MORB samples collected at oceanic ridges exhibit a weak or moderate vesicularity. In Fig.1 are reported vesicularity data obtained by Chavrit (2010) from the analysis of 65 MORB samples collected at the atlantic, pacific and indian ocean ridges and those of Hekinian et al. (2000) and Pineau et al. (2004) for MORB samples dredged on seamounts of the mid-atlantic ridge. These MORB samples were collected between ~1,000 and 5,000 m.b.s.l. (i.e. ~0.1-0.5 kbar of hydrostatic pressure). If one assumes that the observed vesicularity is due to a closed system

310    degassing without loss of vesicles between the source region and the seafloor, then our calculations

311    suggest than most of the MORB samples are issuing from a source region located at shallow depth

312    ($P_{CO_2}^{sat} \leq 5$ kbar i.e. d<15km). Moreover, if exsolution only occurs when some degree of super

313    saturation in $CO_2$ is reached (for illustration, the dotted curves shown in Fig.1 correspond to a degree

314    of super saturation α=2), our calculations indicate that the vesicularity observed at eruption will be

315    significantly lowered if the source region of the $CO_2$-saturated magma is located at shallow depth (for

316    instance, if $P_{CO_2}^{sat}$=1 kbar, the vesicularity at 4,000 m.b.s.l. is three times smaller for a melt exhibiting a

317    super saturation α=2 than for a melt erupting at equilibrium). Incidentally, a super saturation in $CO_2$ is

318    frequently reported in the bulk melt of MORB samples (Dixon et al., 1988; Jendrzejewski et al., 1997;

319    Aubaud et al., 2004; Cartigny et al., 2008) and it seems to be higher when the vesicularity is low

320    (<1%, see Chavrit, 2010).

321        In contrast, for $CO_2$-rich magmas their high vesicularity at eruption is barely affected by a possible

322    super saturation (see dotted curves in Fig.1 when $V^* \geq 10\%$). With regard to those rare MORB

323    samples exhibiting a high vesicularity (>10% in Fig.1), their source region is likely located at a depth

324    between ~15 km and ~90km (i.e. $P_{CO_2}^{sat} \sim$ 5-30 kbar). For instance, the *2πD43 popping rock*, which is

325    considered as an undegassed MORB sample, likely is issuing from a $CO_2$-saturated magma located at

326    ~36 km depth ($P_{CO_2}^{sat}\sim$12 kbar, according to Fig.1). In the same way, the highly vesiculated samples

327    ($V^* \geq 0.50$ in Fig.1) collected in the rift valley of the mid-atlantic ridge near 34°N (Hekinian et al.,

328    2000; Pineau et al., 2004), and which are associated with volcaniclastic deposits, could be issued from

329    $CO_2$-saturated magmas at $P_{CO_2}^{sat}\sim$20-28 kbar (see Fig.1), a source region located about 60-90 km in the

330    oceanic mantle. Moreover, according to our results, a magma saturated in $CO_2$ at 30 kbar reaches the

331    fragmentation limit, $V^*$=0.75 (Sparks, 1978; Namiki and Manga, 2008), if it erupts at about 1,600

332    m.b.s.l., precisely the depth at which volcaniclastic deposits are observed. Notice that to reach the

333    fragmentation limit at a greater water depth (e.g. 4,000 m.b.s.l.) the magma must be saturated in $CO_2$

334    at a much higher pressure (e.g. ~50 kbar or at 150 km in the upper mantle, according to Fig.1).

335    A more quantitative way to compare the vesicularity of MORB samples with that deduced from

336    simulation data is to represent the vesicularity measured (or evaluated) at a given hydrostatic pressure

337    as function of the total $CO_2$ content in the undegassed sample. Experimentally, the $CO_2$ content is

338    evaluated by summing the amount of $CO_2$ in the bulk melt, measured by step heating or by IR

339    spectroscopy, and that in the vesicles, measured by crushing or estimated from the vesicularity in

340    treating the $CO_2$ phase filling the vesicles as an ideal gas (Graham and Sarda, 1991; Javoy and Pineau,

341    1991; Hekinian et al., 2000; Pineau et al., 2004). However, the $CO_2$ content in the glass is generally

342    much smaller than the one in the vesicles, except for samples exhibiting a very low vesicularity

343    ($V^* \leq 10^{-3}$). In Fig.2 is reported the vesicularity as function of the total $CO_2$ content (in $g_{CO_2}/g_{melt}$ ) of

344    MORB samples collected at the Atlantic, Indian and Pacific ocean ridges at various water depth

345    (minimum and maximum water depths are indicated in the figure). These experimental data (Chavrit

346    (2010) and Pineau etal. (2004)) are compared with our calculations for a $CO_2$-saturated magma

347    ascending under isothermal condition (T=1873 K) and erupting at 1,000 m.b.s.l. (upper dots) or at

348    5,000 m.b.s.l. (lower dots), these two depths enclosing approximately the range of depth encountered

349    at oceanic ridges. The agreement with experimental data is remarkable, especially if one emphasizes

350    that the vesicularity of MORBs is evaluated from glassy samples whereas the vesicularity of our

351    model is for a liquid sample erupting at 1873 K. It is commonly admitted (even if this is questionable)

352    that the vesicularity of a magma erupting on the seafloor is frozen in when the temperature of the melt

353    passes through the glass transition temperature ($T_g$~1000 K, Ryan and Sammis (1981)). But if one

354    considers that the solubility of $CO_2$ at the pressure of eruption changes very little when the

355    temperature drops (Pan et al., 1991), and that the $CO_2$ phase filling the vesicles behaves as an ideal

356    gas, then the vesicularity at $T_g$ will be smaller than the vesicularity of the high temperature liquid just

357    before eruption by a ratio equal to $T_g/T$ (~0.53 for the simulated MORB at 1873 K). Nevertheless, this

358    estimation has to be taken with some caution as it likely overestimates the decrease of the vesicularity

359    during the quench. As a matter of fact, upon rapid cooling, the viscosity of a basalt can be sufficiently

360    high even above $T_g$ to limit vesicle shrinkage (Gardner et al., 2000). However, the important point is

361    that a small decrease of the vesicularity is expected in going from the liquid to the glass, a feature which can explain why the calculated vesicularity is systematically (but slightly) larger than the observed one in Fig.2. For instance, the *2πD43 popping rock* dredged at 3,770 m on the mid-atlantic ridge has a vesicularity of 17% and a $CO_2$ content evaluated around 0.85-1.24 wt% (Graham and Sarda, 1991; Javoy and Pineau, 1991; Cartigny et al., 2008; Chavrit, 2010) as compared with 20-28% for the vesicularity of our simulated melt at 1873 K bearing the same $CO_2$ content. It is noteworthy that the calculations presented in Fig.2 do not take into account a possible super saturation in $CO_2$. We have emphasized earlier that the vesicularity of a melt erupting on the seafloor is significantly affected by super saturation only when the source region is very shallow (a few km in the oceanic crust). In this case the vesicularity of the super saturated melt can be much smaller than the one of a sample at equilibrium (this is illustrated in Fig.1 for a super saturation equal to 2). The very low vesicularity exhibited by MORB samples of the Pacific ridge (see the crosses in Fig.1 and 2) can then be explained in noting that their degree of super saturation is in the range 1.5-2.5 (Chavrit, 2010).

In conclusion, the large distribution of vesicularity which characterizes MORB samples can be described by a $CO_2$ degassing process initiated at various depths in the oceanic mantle. However, the vesicularity oberved after eruption on the seafloor is only the evidence of the last vesiculation episode experienced by the ascending magma, an evidence which gives no information about the degassing history prior this episode (if any). But, as discussed in the following, complementary information about the degassing trajectory can be obtained from the inventory of noble gases.

### 3.2 Partitioning and fractionation of noble gases.

According to Eqs.(4) and (5), the partitioning of noble gases between melt and vesicles is controlled by $\gamma_m^i$ and $\gamma_v^i$, the noble gas solubility parameters in the two coexisting phases. Values of these parameters evaluated by the TPM at isothermal condition (T=1873 K) are listed in Table 2 as function of the pressure. For a given noble gas (*He*, *Ne*, *Ar* or *Xe*) both solubility parameters, $\gamma_m^i$ and $\gamma_v^i$, decrease strongly when the pressure is increased, whereas at a given pressure, the heavier the noble gas the smaller the solubility parameters in the two coexisting phases ($\gamma^{He} > \gamma^{Ne} > \gamma^{Ar} > \gamma^{Xe}$). These trends are similar to those obtained by Guillot and Sator (2012) in their study of the solubility of noble 388    gases in dry silicate melts. According to Eq.(2) the ratio, $\gamma_m^i/\gamma_v^i$, is equal to , $\rho_m^i/\rho_v^i$, where $\rho_m^i$ is the

389    number density of noble gas of species $i$ in the $CO_2$-saturated melt (i.e. the number of noble gas atoms

390    in the melt per volume of melt) and $\rho_v^i$ is the one in the $CO_2$ phase. This ratio is nothing but the noble

391    gas partition coefficient which expresses the distribution of the noble gas ($i$) between the two phases.

392    As shown in Fig.3, the noble gas partition coefficient increases with the pressure but tends to level off

393    above 50 kbar. More precisely, between 0 and 50 kbar the partition coefficient for $He$ increases by

394    ~400% , that for $Ne$ by ~500% and that for $Ar$ by ~600%. This pressure effect implies that the transfer

395    of a noble gas from the melt to the vesicles is hindered at depth (high P) by the high density of the $CO_2$

396    phase filling the vesicles. So, to properly estimate the melt/$CO_2$ partitioning of noble gases in a magma

397    at depth, it is inaccurate to use the low pressure solubility of noble gases in the basaltic melt and to

398    consider the $CO_2$ phase as an ideal gas (for which $\gamma_v^i =1$).

399    The inventory of noble gases in MORB samples is well documented (Graham, 2002). In the

400    following our analysis will focus on $^4He$ and $^{40}Ar$ for which a consistent set of data is available. Before

401    to compare these data with the prediction of our model calculation, it is worthwhile to examine how

402    $He$ and $Ar$ distribute between melt and vesicles for a closed system degassing. In Fig.4 is shown the

403    evolution of $He$ and $Ar$ contents, as calculated from Eqs.(4) and (5), in melt and in vesicles as function

404    of the vesicularity evaluated for an hydrostatic pressure equivalent to 3,000 m.b.s.l. (the average water

405    depth at oceanic ridges). The $He$ and $Ar$ contents are normalized with their initial abundance in the

406    undegassed melt. Various initial $CO_2$ contents are investigated (controlled by the parameter $W_0$ in

407    Eq.(7)). They correspond to a $CO_2$-saturation pressure, $P_{CO_2}^{sat}$, varying from 100 to 0.4 kbar (the value

408    of $P_{CO_2}^{sat}$ is indicated in Fig.4). An attentive examination of Fig.4 shows that when the vesicularity is

409    large (i.e. $V^* \geq 10\%$ or $P_{CO_2}^{sat}>5$ kbar), more than 99% of $Ar$ atoms and 90% of $He$ atoms are in the

410    vesicles at eruption, whereas for a 1% vesicularity, about 90% of $Ar$ atoms are in the vesicles but only

411    40% of $He$ atoms. If outgassing starts at very shallow depth, for instance at 1,000 m below the seafloor

412    (i.e. at $P_{CO_2}^{sat}=0.4$ kbar), the vesicularity becomes as small as 0.2% and yet 60% of $Ar$ atoms are in the

413    vesicles when 90% of $He$ atoms are staying in the melt. These results are easily understandable if one recalls that the melt/$CO_2$ partition coefficient for *He* is much greater than for *Ar* (a factor of ~10, see Fig.3). So, according to the vesicularity of the sample at eruption, the fractionation of noble gases (i.e. the ratio *He/Ar*) in melt and in vesicles will be different.

In order to compare in absolute value the noble gas contents measured in MORB samples with those calculated from Eqs.(4) and (5), we make the assumption that the concentration in $^{40}Ar$ and the $^4He/^{40}Ar$ ratio extracted from the *2πD43 popping rock* reflect best the noble gas composition of the source region for MORBs (~0.1 ppmw of $^{40}Ar$ after correction for atmospheric contamination, and $^4He/^{40}Ar$ ~1.5, see Moreira et al. (1998), Moreira and Sarda (2000) and Raquin et al. (2008)). In Fig.5 is presented the evolution of *Ar* content in vesicles (in $10^{-6}$ $g_{Ar}/g_{melt}$) as function of the vesicularity of the simulated MORB evaluated at 3,000 m.b.s.l., and in Fig.6 is presented the evolution of the $^4He/^{40}Ar$ ratio. Here again, several initial conditions in $CO_2$ content for the source region are considered (the corresponding $CO_2$-saturation pressure evolving from 50 to 0.4 kbar). In both figures are shown literature data (black squares) compiled by Chavrit (2010) and obtained by crushing MORB samples (for this reason only noble gas contents in vesicles are reported). It is clear that a simple closed system degassing (see the red dots defining the curve with index $V^1$ in Figs.5 and 6) is unable to reproduce the literature data except those of the *2πD43 popping rock* which are well represented by a single stage of vesiculation starting at ~30 km depth ($P_{CO_2}^{sat}$~10 kbar). For the other MORB samples, the *Ar* contents in vesicles are found to be much weaker (by one to three orders of magnitude) than those predicted by a one-stage vesiculation mechanism. Correlatively, the $^4He/^{40}Ar$ ratios observed in MORB samples are generally much higher than predicted.

These findings imply that either the initial noble gas contents of the MORB source vary from one sample to another, or the degassing trajectory is more complex than the one described by a simple closed system degassing. Although it cannot be excluded that for some settings the noble gas contents may vary significantly or be affected by diffusive fractionation during mantle melting (Burnard, 2004; Burnard et al., 2004; Yamamoto et al., 2009), it is generally accepted that the upper mantle is compositionally homogeneous with regard to noble gases, and because the latter ones are incompatible

440    elements (Brooker et al., 1998; Chamorro et al., 2002; Brooker et al., 2003; Heber et al., 2007), the

441    melt composition in noble gases reflects that of the solid mantle. In contrast, the C contents of the

442    upper mantle likely is more heterogeneous (Saal et al., 2002; Cartigny et al., 2008). Furthermore it

443    may happen that during ascent the magma loses its $CO_2$-rich bubbles (e.g. slow ascent in a tortuous

444    conduit),with the noble gases dissolved into it. Next, after this resetting step, the ascending melt starts

445    again to degas when the residual amount of $CO_2$ oversteps the saturation limit. If this stage of vesicle

446    loss occurs when the vesicularity of the magma is significant (a few percent or more) then the residual

447    melt is strongly depleted in noble gases as a large proportion of the latter ones have been taken away

448    with the $CO_2$ bubbles (see Fig.4 for the respective proportion of noble gases in melt and in vesicles).

449    According to Sarda and Moreira (2002), Sarda and Guillot (2005) and Guillot and Sarda (2006),

450    several stages of vesiculation followed by vesicle loss can explain the low $^{40}Ar$ contents and the high

451    $^{4}He/^{40}Ar$ ratio measured in many of MORB samples.

452        In Figs.5 and 6 are reported the $Ar$ content and the $^{4}He/^{40}Ar$ ratio in vesicles of the simulated MORB

453    after having experienced several stages of vesiculation with vesicle loss. In practice, after each stage of

454    vesicle loss, the noble gas content, $N_m^i$ (with $i = He$ or $Ar$), figuring in Eq.(4), is substituted to $N_0^i$ in

455    Eqs.(4) and (5) while $V^*$ is set to 0. For illustration, in Figs.5 and 6 are shown 2-stage (marked $\mathbf{V^2}$), 3-

456    stage (marked $\mathbf{V^3}$) and 4-stage (marked $\mathbf{V^4}$) vesiculations. In the case of $\mathbf{V^2}$, three degassing paths are

457    presented (see blue dots): ($\mathbf{P_1}$=20; $\mathbf{P_2}$=19, 18,.., 0.4), ($\mathbf{P_1}$=5; $\mathbf{P_2}$=4, 3,.., 0.4) and ($\mathbf{P_1}$=1; $\mathbf{P_2}$=0.9, 0.8,..,

458    0.4). The first number $\mathbf{P_1}$ corresponds to the $CO_2$ saturation pressure (in kbar) at which the first

459    vesiculation stage occurs, and the second number $\mathbf{P_2}$ is the saturation pressure of the second

460    vesiculation stage in assuming that in between the two stages, the ascending MORB melt has

461    experienced a total loss of vesicles. In looking at Fig.5, one notices that the higher the pressure

462    threshold $\mathbf{P_1}$ of the first vesiculation (i.e. the deeper in the oceanic mantle the first degassing episode),

463    the lower the $Ar$ contents in vesicles of a MORB sample of a given vesicularity at eruption (e.g. for

464    $V^*$=1% i.e. for $\mathbf{P_2}$=0.8 kbar, the $Ar$ content is equal to ~0.04 ppmw if $\mathbf{P_1}$=1 kbar, ~0.0035 ppmw if

465    $\mathbf{P_1}$=5 kbar and ~0.0006 ppmw if $\mathbf{P_1}$=20 kbar). As shown in Fig.5, it is clear that a 2-stage vesiculation

466    process may account of a significant part of the literature data (compare the squares with the blue dots). However, 3-stage and 4-stage vesiculations (or higher order) may be necessary to describe highly *Ar*-depleted samples (<10⁻⁴ ppmw): two exemples are given for illustration, $\mathbf{V^3}$ ($\mathbf{P_1}$=20; $\mathbf{P_2}$= 3; $\mathbf{P_3}$=2, 1,.., 0.4 see green dots) and $\mathbf{V^4}$ ($\mathbf{P_1}$=20; $\mathbf{P_2}$= 3; $\mathbf{P_3}$= 1; $\mathbf{P_4}$= 0.9,..,0.4 see cyan dots). Incidentally, one notices that most of the measured vesicularities are below ~5%, a value which suggests that most of MORB samples have experienced a last vesiculation episode not as deep as ~10 km in the mantle (or $P_{CO_2}^{sat}$ <3kbar).

With regard to the $^4He/^{40}Ar$ ratio presented in Fig.6, similar conclusions are obtained (compare squares with dots in the figure). Thus in introducing a 2-stage vesiculation the $^4He/^{40}Ar$ ratio lies between 1 and 10, whereas higher values (~10-100) are reached with 3- or 4-stage vesiculations. Nevertheless, the highest values of the $^4He/^{40}Ar$ ratio (in the range 100-1000) and the lowest Ar contents (<10⁻⁴ ppmw) measured in MORB could originate from a Rayleigh distillation process (Moreira and Sarda, 2000; Burnard, 2001) where $CO_2$ bubbles are extracted from the melt as soon as they are formed (e.g. in a shallow magma chamber). But other mechanisms have been proposed to explain high values of the $^4He/^{40}Ar$ ratio as, for instance, a fractional crystallization-assimilation-degassing in a shallow magma chamber (Marty and Zimmermann, 1999) or a kinetic fractionation controlled by diffusion during incomplete degassing (Aubaud et al., 2004; Paonita and Martelli, 2006, 2007). The cornerstone of this latter mechanism is that the larger diffusivity of *He* atoms into the MORB melt with respect to that of *Ar* and $CO_2$ may enrich in *He* atoms incipient $CO_2$ bubbles, a kinetic disequilibrium which can be frozen in at eruption if the ascent rate is sufficiently high. We show in the next section that this hypothesis can be tested by MD simulation.

## 4. Kinetics of noble gas transfer during bubble formation

In the course of a MD simulation it is possible to reproduce a degassing process in applying a pressure drop to a volatile-saturated melt. For instance, in starting from a well equilibrated $CO_2$ bearing melt at 1873K and 50 kbar (the $CO_2$ content is ~ 9.5 wt%) and in applying a sudden pressure drop from 50 to 5 kbar, a 3D visualization of the simulation box shows the growth of a $CO_2$ nanobubble fully developed after about a few ns of running time (see Fig.7). The vesicularity of the

493 MORB sample after the pressure drop is that expected from the thermodynamic relationship (see

494 Eq.(7) and Fig.1), i.e. about 30 % in the present case. Correspondingly the bulk melt around the $CO_2$

495 bubble has a $CO_2$ content of ~0.3 wt%, a value in agreement with the solubility of $CO_2$ at these

496 thermodynamic conditions (in Fig.7, the initial state is a melt made of 5,000 ions, ~176 carbonate ions

497 and ~99 $CO_2$ molecules, whereas the final state is a melt composed of 5,000 ions, ~2 carbonate ions

498 and ~6 $CO_2$ molecules and a bubble formed by ~267 $CO_2$ molecules). We have repeated the above

499 degassing experiment for other thermodynamic conditions (e.g. 100→50 kbar, 100→10 kbar, and

500 30→3 kbar) and the final state is characterized by a single $CO_2$ bubble whose the size agrees with the

501 vesicularity estimated by thermodynamics. But, because of the small system size investigated (a few

502 thousand particles), only a few nucleation loci are observed initially in the simulation box (see Fig.7),

503 and hence it is not possible to investigate the nucleation rate quantitatively (very large system sizes are

504 then required). In practice, to render more effective bubble nucleation it is important to deal with a

505 high initial $CO_2$ content, i.e. to prepare the initial state at a high enough pressure (e.g. 100, 50 or 30

506 kbar), and to apply a large decompression rate (here the pressure drop is applied quasi instantaneously

507 over one MD step, i.e. $10^{-15}$ s, a more detailed analysis of bubble nucleation by changing the

508 decompression rate is beyond the scope of this article). But this is immaterial with regard to our

509 primary objective which is to quantify how fast is the transfer of noble gases from the melt to a

510 growing $CO_2$ bubble.

511 A small number (~10) of *Ar* or *He* atoms is incorporated into a $CO_2$-saturated MORB melt at given

512 (T,P) conditions, and after a period of equilibration (~1 ns) the system is suddenly decompressed (e.g.

513 from 100→10 kbar). It may happen that the number of noble gas atoms incorporated into the melt is

514 high enough for the melt to be supersaturated in noble gases (this is the case for Ar atoms but not for

515 He atoms in the pressure range investigated, for a discussion on noble gas solubility in silicate melts

516 see Guillot and Sator (2012)). However there is no evidence of noble gas atom clustering in the

517 simulated melt before decompression, likely because the nucleation rate for such an event is so low

518 that the time to spend for observing noble gas nucleation exceeds by far our simulation time. Thus

519 during the $CO_2$ degassing process, each noble gas atom behaves independently from each other and for

520  this reason they are used as independent tracers. We have checked this point by performing simulation

521  runs with a single noble gas atom and have found essentially the same results (except a poorer

522  statistics).

523  In order to follow the kinetics of bubble formation, we have evaluated as function of running time

524  the average elemental fraction (on an atomic basis), $f_{CC}$, of $CO_2$ molecules which are in the immediate

525  vicinity (first shell) of a carbon atom (this one belonging to a $CO_2$ molecule or to a $CO_3^{2-}$ carbonate

526  ion, without distinction). The radius ($r_s$=5A) of the first shell around each carbon atom was chosen

527  empirically from an analysis of the atom-atom pair distribution functions, and we have checked that

528  the results are robust when changing its value on a limited range. During a decompression experiment

529  $f_{CC}$ is expected to vary from a low or moderate value to a high value. Indeed, before decompression

530  each carbon atom of the $CO_2$-saturated melt is surrounded primarily by the elements of the silicate,

531  whereas after the pressure drop a majority of $CO_2$ molecules are in the bubble.

532  To improve the statistics, our results were averaged over a series of 20 independent decompression

533  experiments (each one long of 5ns). For illustration, the time evolution of $f_{CC}$ is shown in Fig.8 for a

534  100→10 kbar pressure drop. Before the pressure drop the value of $f_{CC}$ in the $CO_2$-saturated melt at

535  equilibrium is ~0.40 (a rather high value because the $CO_2$ content is ~29wt% at 100 kbar), and after

536  the pressure drop $f_{CC}$ increases rapidly to reach a plateau value, ~0.76, after about 2,000 ps. This

537  evolution of $f_{CC}$ is the signature of $CO_2$ bubble formation as shown by the correlated evolution of the

538  vesicularity of the simulated sample (see V* in Fig.8). Indeed, when a bubble nucleates and growths,

539  the silicate atoms surrounding a carbon atom are progressively replaced by $CO_2$ molecules. Notice that

540  in a bulk $CO_2$ fluid $f_{CC}$ is equal to one, but in our simulations the high surface/volume ratio of the

541  nanobubble modifies this value (hence $f_{CC}$<1). As a matter of fact, the $CO_2$ molecules in contact with

542  the melt at the surface of the bubble represent a significant proportion of the bubble forming molecules

543  (~40% for a bubble of 40A in diameter).

544  A quick glance at Fig.8 reveals that the time evolution of $f_{CC}$ (and V* as well) exhibits two different

545  regimes: a rapid evolution of great amplitude at short times (0<t<100 ps) followed by a slower

546 evolution of a much weaker amplitude at long times (100<t<2,000 ps). This time evolution is well
547 described by the following bi-exponential function,

548 $$f_{CC} = \left(f_{CC}^{min} - f_{CC}^{max}\right)e^{-t/\tau_1} + f_{CC}^{inter}\left(e^{-t/\tau_2} - e^{-t/\tau_1}\right) + f_{CC}^{max} \tag{10}$$

549 where $f_{CC}^{min}$, $f_{CC}^{inter}$ and $f_{CC}^{max}$ are fitting parameters and $\tau_1$ and $\tau_2$ are time constants (the curve fitting
550 shown in Fig.8 corresponds to the following parameters: $f_{CC}^{min} = 0.40$, $f_{CC}^{inter} = 0.04$, $f_{CC}^{max} = 0.76$,
551 $\tau_1 = 60\ ps$ and $\tau_2 = 600\ ps$). The two time constants give an estimate of the bubble growth rate at
552 the nanometric scale. The time constant $\tau_1$, associated with the short time evolution, corresponds to
553 the time spent for the formation of a critical nucleus, i.e. a cluster in which a central $CO_2$ molecule has
554 essentially $CO_2$ molecules as first neighbors. On the other hand, the time constant $\tau_2$ gives an estimate
555 of bubble growth at long time. It expresses the coalescence time of critical nuclei and the expansion
556 duration of bubble growth. Notice that the bubble ceases to grow after about 2,000 ps, when the melt
557 super saturation in $CO_2$ molecules is exhausted, i.e. when the chemical equilibrium between the
558 silicate melt and the $CO_2$ phase is achieved. Consequently the size of the bubble depends on the size of
559 the simulated sample but the vesicularity does not.

560 Although our simulation gives information on bubble formation at the nanometric scale, for the sake
561 of completeness it is useful to compare this microscopic description with a macroscopic point of view
562 as given by classical nucleation theory. In this framework, for early-stage bubble growth when the
563 surface/volume ratio is large, growth is limited by the viscosity of the melt (Prousevitch et al., 1993;
564 Toramaru, 1995; Navon et al., 1998), and the bubble radius grows exponentially with time,

565 $$R(t) \approx \exp\left[\frac{\Delta P}{4\eta}t\right] \tag{11}$$

566 where $\Delta P$ is the difference between the volatile pressure inside the bubble and the ambient pressure
567 (equated to the pressure drop in a first approximation) and $\eta$ the melt viscosity. In our simulation, the
568 volume of an initial nucleus being proportional to the number of incoming $CO_2$ molecules, the latter

569 one grows approximately as $R^3(t)$. From the analogy between the microscopic definition expressed by

570 Eq.(10) and the macroscopic relationship given by Eq.(11) one can deduce the following relation,

571 $$\tau_1 \sim \frac{4\eta}{3\Delta P} \qquad\qquad\qquad (12)$$

572 where $\tau_1$ is the formation time of a critical nucleus. For the decompression experiment presented in

573 Fig.8, the Eq.(12) leads to $\tau_1 = 6.7$ ps in using $\Delta P = 90$ kbar and $\eta = 0.045$ Pa.s for the viscosity of our

574 $CO_2$-bearing MORB melt evaluated at 1873K and 10 kbar (notice that the viscosity of our simulated

575 melt is much lower than that of a real basaltic melt, by roughly a factor of ten at 1873 K, for a

576 discussion see Bauchy et al. (2012), but this is immaterial in the present context). But this value of $\tau_1$

577 is about 9 times shorter than the result of the simulation ($\tau_1^{sim}= 60$ ps), a finding which implies that the

578 simulation and the nucleation theory do not probe the same process or that the strict application of

579 Eq.(11) is doubtful over a time scale of a few ps (for instance the relaxation time for the viscosity is

580 much larger than a ps). Moreover the Eq.(11) is valid so long as a well formed interface between the

581 $CO_2$ bubble and the silicate melt can be identified, and this is not yet the case when the system is made

582 of scattered clusters. Nevertheless, for nucleation theory, the scale of the critical nucleus radius (given

583 by $2\gamma/P$, where $\gamma \sim 0.36$N/m is the surface tension for molten basalt as evaluated by Walker and

584 Mullins, 1981), is about 3.8A at P=10 kbar, a value which agrees well with the size of clusters

585 observed in our simulation during the early stage of bubble formation. Moreover, the nucleation time

586 $\tau_1$ increases strongly when the volatile initial concentration becomes lower or when the pressure drop

587 is decreased (e.g. $\tau_1$ evolves from $\sim60$ ps to $\sim230$ ps when $\Delta P$ varies from 90 to 50 kbar). All these

588 findings are in a qualitative agreement with classical nucleation theory (Proussevitch and Sahagian,

589 1998) which predicts a slowing down of bubble growth at once with lower volatile oversaturation and

590 with lower initial volatile concentration.

591 For noble gases we have proceeded similarly by defining the average fractions in $CO_2$

592 molecules, $f_{HeC}$ and $f_{ArC}$, within a sphere of radius 5A centered on a $He$ or an $Ar$ atom. As illustrated

593 in Fig.8, an important result is that the time evolutions of $f_{HeC}$ and $f_{ArC}$ are not only very similar to 594    each other but they are virtually identical to that of $f_{CC}$. More precisely they can be described by

595    Eq.(10) with $\tau_1^{He} \approx \tau_1^{Ar} (\approx \tau_1^C) = 60\ ps$, $\tau_2^{He} \approx \tau_2^{Ar} (\approx \tau_2^C) = 600\ ps$, $f_{HeC}^{min} = 0.42$, $f_{HeC}^{inter} = 0.04$,

596    $f_{HeC}^{max} = 0.78$, $f_{ArC}^{min} = 0.50$, $f_{ArC}^{inter} = 0.04$, and $f_{ArC}^{max} = 0.83$. Moreover, there is no time lag between

597    the rises of $f_{HeC}$ and $f_{ArC}$ and that of $f_{CC}$, they are concomitant with each other. In other words, as

598    soon as nucleation takes place, the *He* and *Ar* atoms located nearby a nucleation locus attach to it and

599    become observers of its evolution. At the scale of our simulation (the simulation cell is a few nm long)

600    the transfer of *He* and *Ar* atoms from the melt to a $CO_2$ bubble is achieved before the latter one is fully

601    developed. There is no quantitative difference between *He* and *Ar* even if *He* atoms are more mobile

602    than *Ar* atoms in the melt (at 1873K and 10 kbar their respective diffusion coefficients differ by

603    roughly one order of magnitude: $D_{He} \sim 10^{-8}$ m$^2$/s and $D_{Ar} \sim 1.5\ 10^{-9}$ m$^2$/s after our own evaluation). In

604    fact the limiting step is the nucleation process itself and the noble gases present in the halo

605    surrounding a nucleation locus stick to it and works one's way into it. A 3D visualization of the atomic

606    trajectories into the simulation box is fully illustrative on this point (see Fig.7).

607    After the formation of the nanobubble is completed (equilibrium), it is observed along the

608    simulation run that the rapid diffusion of a *He* atom into the $CO_2$ bubble leads from time to time to

609    brief excursions of the noble gas atom into the bulk melt. This behavior is expected because the

610    melt/$CO_2$ partition coefficient of a *He* atom being equal to about 3% at 5 kbar (see Fig.3), in a highly

611    vesiculated basaltic melt ($V^* \sim 0.3$ in the exemple of Fig.7) the noble gas will spent approximately

612    ~3% of its time in the bulk melt and ~97% in the $CO_2$ bubble. For *Ar* atoms the partition coefficient is

613    ten times smaller (~0.3% at 5 kbar, see Fig.3) and indeed we observe very scarcely an excursion of an

614    *Ar* atom out of the bubble. In summary, during a degassing episode the noble gas atoms present in the

615    halo surrounding an incipient $CO_2$ bubble are transferred in the latter one at the same rate than the $CO_2$

616    molecules themselves. So, when bubble growth is ended (or stopped) the concentration in noble gases

617    becomes stationary and its value in the vesicles can be evaluated from Eq.(5) in introducing for $V^*$ the

618    current vesicularity.

619     Although our calculations were performed at the nanometric scale we believe that the conclusions

620    which can be inferred are meaningful at the macroscopic scale. Thus, our results don't support the

621    hypothesis that a kinetic fractionation between *He* and *Ar* is at the origin of the high *He/Ar* ratio

622    observed in vesicles of some MORB samples (Aubaud et al., 2004). According to this hypothesis, the

623    transfer of noble gases from the melt to the bubbles is considered as decoupled from the very process

624    of bubble formation. When degassing is taking place, the *He* atoms diffusing faster than *Ar* atoms into

625    the melt, the $CO_2$ bubbles (assumed to be depleted in noble gases) would be transiently enriched in *He*

626    atoms with respect to *Ar* atoms, a disequilibrium leading to increase the *He/Ar* ratio in the vesicles.

627    Furthermore, if the magma erupts before degassing is completed (due to a rapid ascent rate for

628    instance) the high *He/Ar* ratio can be imprinted in the MORB glass (Paonita and Martelli, 2006). Our

629    results show that the prerequisite of this scenario (the decoupling between bubble formation and noble

630    gas transfer) is unrealistic. As a matter of fact, the noble gases are present from the formation of

631    critical nuclei and consequently their abundances in the vesicles are controlled by the current

632    vesicularity of the sample. For instance, a MORB melt $CO_2$-saturated at 0.8 kbar and erupting at 3,000

633    m.b.s.l. (i.e. $P_{erup}$~0.3 kbar) will exhibit at equilibrium a vesicularity about 1% (see Fig.1). In this case

634    the *He/Ar* ratio in the vesicles (see Fig.4) is equal to ~$0.4(C_0^{He}/C_0^{Ar})$ where $C_0^{He}$ and $C_0^{Ar}$ are the noble

635    gas contents before vesiculation (notice that $C_0^{He}/C_0^{Ar}$ may vary with the source region, see the

636    discussion in section 3.2). But if during magma ascent, the $CO_2$ exsolution is delayed up to a super

637    saturation ratio $\alpha$=2.6 is reached (for exemple), then the vesicularity at eruption is as low as ~1.5 $10^{-4}$

638    and the corresponding *He/Ar* ratio is equal to ~$0.1(C_0^{He}/C_0^{Ar})$. Thus, for a closed system degassing, the

639    higher the $CO_2$ super saturation ratio the lower the *He/Ar* ratio in the vesicles at a given pressure

640    (because *He* is more soluble than *Ar* in the silicate melt). Even if some *He* and *Ar* atoms of the melt

641    are sufficiently distant from vesicles for migrating with delay in the latter ones, the incoming fluxes in

642    noble gases are compensated by outgoing fluxes to fulfill equilibrium (Fick's law). In summary, an

643    incomplete degassing lowers the vesicularity (with respect to a complete degassing) and diminishes

644    the *He/Ar* ratio in the vesicles. So, to reconcile theory and observation (MORB samples exhibiting a

645    low vesicularity and a strong $CO_2$ super saturation are generally displaying a high *He/Ar* ratio in the 646 vesicles) one must put forward other mechanisms capable of producing high $C_0^{He}/C_0^{Ar}$ ratio in the melt

647 before the last vesiculation episode. We think that the sequential degassing mechanism with vesicle

648 loss, presented in section 3.2, can solve this outstanding question.

**5. Conclusions**

650 In this paper it has been shown that several aspects of MORB degassing can be described

651 realistically by atomistic simulations. The main results of this study are the following.

652 1- In using a force field developed by Guillot and Sator (2011) to model by MD simulation the

653 incorporation of $CO_2$ into silicate melts, the vesicularity of a MORB melt has been evaluated as

654 function of initial $CO_2$ contents and pressure at eruption. An excellent agreement is obtained between

655 calculated values and vesicularity data of MORB samples collected at mid-ocean ridges. This

656 comparison also indicates that most of MORB samples have experienced a last vesiculation episode at

657 shallow depth (d<15 km) in the oceanic mantle. In contrast, a few highly vesicular samples (e.g.

658 *popping rocks*) could be issued from deep-seated $CO_2$-rich magmas. A possible origin of the latter

659 ones are $CO_2$-rich silica-undersaturated melts (e.g. carbonatite-like) in transit through the oceanic

660 mantle which, in assimilating the silicate matrix, could drive the melt to a basaltic composition. If their

661 $CO_2$ content is sufficiently high, the MORB melt may exsolve a great amount of $CO_2$ leading in

662 special cases to a submarine explosive volcanism. For instance, according to our results, the

663 vesicularity of a MORB composition with 9.5 wt% $CO_2$ reaches the fragmentation threshold

664 ($V^*\sim0.75$) at a water depth of 4,000 m whereas a MORB melt with 4.6 wt% $CO_2$ will experience an

665 explosive fragmentation in erupting at 1,500 m.b.s.l.. In short, the possibility that $CO_2$-rich magmas

666 may drive explosive volcanism at some mid-ocean ridges has been underestimated up to now.

667 2- The evaluation of the pressure dependence of the noble gas partition coefficients between a MORB

668 melt and a supercritical $CO_2$ phase shows that noble gas transfer from the silicate melt to the $CO_2$

669 phase is hindered at high pressure by the density of the fluid. Thus, the partition coefficients of $He$ and

670 $Ar$ increase by a factor of about 4~5 between 0 and 50 kbar. So MORB degassing at great depth would 671    be less efficient than expected to deplete the upper mantle in noble gases. Furthermore it is shown that

672    the large distribution in Ar contents and of the He/Ar ratio measured in MORB samples can be

673    explained if the ascending magma experiences a series of vesiculation at depth followed by vesicle

674    loss.

675    3- By performing a decompression numerical experiment where a pressure drop is applied to a volatile

676    ($CO_2$+noble gas) bearing MORB melt, it is possible to investigate (and to visualize) the formation of

677    critical nuclei leading to $CO_2$ bubble formation. Interestingly, noble gases present in the neighborhood

678    of nucleation loci exhibit a great affinity for $CO_2$ clusters. In fact, the transfer of noble gases from the

679    melt to these embryos is so effective that the transfer rate in a growing $CO_2$ bubble is identical to that

680    of the $CO_2$ molecules themselves. Thus, when bubble growth is stopping the concentration of vesicles

681    in noble gases becomes stationary and is a function of the current vesicularity (see Eq.(4)). This

682    finding doesn't support the hypothesis that a kinetic fractionation between *He* and *Ar* (induced by their

683    diffusivity difference in the melt) is at the origin of the high *He/Ar* ratio observed in vesicles of some

684    MORB samples.

685

686    **Acknowledgements**

687    Bertrand Guillot and Nicolas Sator acknowledge the Agence Nationale pour la Recherche (under

688    Grant agreement ANR-2010-BLAN-621-03) and the European Research Council (under Grant

689    agreement N°279790).

690

691

692

693

694

695

696

697

698

699 **Table 1**

700 Simulation conditions. All simulations were performed at 1873K. In the table below, the simulated
701 MORB melt is composed of 5,000 ions plus the number of $CO_2$ molecules, $(N_{CO_2})_{melt}$, necessary to
702 reach $CO_2$-saturation (indicated by $W$) at the investigated pressure. The density of the $CO_2$-saturated
703 MORB is given by $n_{melt}$ and that of the coexisting supercritical $CO_2$ phase by $n_{CO_2}$. To evaluate the
704 system size dependence, the calculations were also performed with a smaller system where the MORB
705 was composed of only 1,000 ions plus the corresponding number of $CO_2$ molecules (in that case
706 $(N_{CO_2})_{melt}$ reported below has to be divided by 5). As for the pure $CO_2$ phase, it was also simulated
707 with two system sizes, a large system with 2,500 $CO_2$ molecules and a smaller one with 500
708 molecules.

| | P(kbar) | $W(g_{CO_2}/g_{melt})$ | $(N_{CO_2})_{melt}$ | $n_{melt}$(g/cm$^3$) | $n_{CO_2}$(g/cm$^3$) |
|---|---|---|---|---|---|
| 709 | | | | | |
| 710 | 0.001 | $5.10^{-7}$ | 0 | 2.63 | $3.10^{-4}$ |
| 711 | 10 | 0.006 | 15 | 2.74 | 0.94 |
| 712 | 20 | 0.016 | 40 | 2.83 | 1.21 |
| 713 | 30 | 0.046 | 120 | 2.88 | 1.38 |
| 714 | 50 | 0.095 | 275 | 2.99 | 1.60 |
| 715 | 80 | 0.21 | 660 | 3.07 | 1.81 |
| 716 | 100 | 0.29 | 1065 | 3.09 | 1.91 |

717

718

719

720

721

**Table 2**

723  Pressure dependence of noble gas solubility parameters in the $CO_2$-saturated MORB melt ($\gamma_m^i$) and in

724  the coexisting supercritical $CO_2$ phase ($\gamma_{CO_2}^i$) at 1873K. Values calculated with the two system sizes

725  are indicated: S is for small system sizes and L is for large system sizes (see Table 1 for the number of

726  atoms in each system). For each investigated pressure, the reference value of the solubility parameter

727  (noted *mean*) is given by the arithmetic mean of the S and L values. Notice that the evaluation of $\gamma_m^{Xe}$

728  becomes inaccurate from 30 kbar whereas that of $\gamma_{CO_2}^{Xe}$ is accurate up to 80 kbar.

| P(kbar) | | $\gamma_m^{He}$ | $\gamma_{CO_2}^{He}$ | $\gamma_m^{Ne}$ | $\gamma_{CO_2}^{Ne}$ | $\gamma_m^{Ar}$ | $\gamma_{CO_2}^{Ar}$ | $\gamma_m^{Xe}$ | $\gamma_{CO_2}^{Xe}$ |
|---|---|---|---|---|---|---|---|---|---|
| 0.001 | S | 1.83(-2) | 1.0 | 9.81(-3) | 1.0 | 1.47(-3) | 1.0 | 2.88(-4) | 1.0 |
| | L | 1.83(-2) | 1.0 | 9.81(-3) | 1.0 | 1.47(-3) | 1.0 | 2.88(-4) | 1.0 |
| | *mean* | *1.83(-2)* | *1.0* | *9.81(-3)* | *1.0* | *1.47(-3)* | *1.0* | *2.88(-4)* | *1.0* |
| 10 | S | 5.49(-3) | 1.71(-1) | 2.87(-3) | 1.39(-1) | 1.96(-4) | 6.60(-2) | 1.39(-5) | 3.21(-2) |
| | L | 5.56(-3) | 1.72(-1) | 2.97(-3) | 1.39(-1) | 2.18(-4) | 6.67(-2) | 1.40(-5) | 3.26(-2) |
| | *mean* | *5.53(-3)* | *1.72(-1)* | *2.92(-3)* | *1.39(-1)* | *2.07(-4)* | *6.64(-2)* | *1.40(-5)* | *3.24(-2)* |
| 20 | S | 2.81(-3) | 6.17(-2) | 1.25(-3) | 4.23(-2) | 4.50(-5) | 1.03(-2) | 1.20(-6) | 2.59(-3) |
| | L | 2.87(-3) | 6.18(-2) | 1.28(-3) | 4.24(-2) | 4.92(-5) | 1.04(-2) | 1.26(-6) | 2.62(-3) |
| | *mean* | *2.84(-3)* | *6.18(-2)* | *1.27(-3)* | *4.24(-2)* | *4.71(-5)* | *1.04(-2)* | *1.23(-6)* | *2.61(-3)* |
| 30 | S | 1.54(-3) | 2.72(-2) | 5.91(-4) | 1.60(-2) | 1.14(-5) | 2.11(-3) | 9.31(-8) | 2.91(-4) |
| | L | 1.59(-3) | 2.70(-2) | 6.12(-4) | 1.59(-2) | 1.35(-5) | 2.11(-3) | 3.24(-7) | 2.92(-4) |
| | *mean* | *1.57(-3)* | *2.71(-2)* | *6.02(-4)* | *1.60(-2)* | *1.25(-5)* | *2.11(-3)* | *2.09(-7)* | *2.92(-4)* |
| 50 | S | 5.21(-4) | 6.66(-3) | 1.49(-4) | 2.97(-3) | 9.94(-7) | 1.27(-4) | 1.05(-8) | 5.92(-6) |
| | L | 5.31(-4) | 6.78(-3) | 1.54(-4) | 3.04(-3) | 1.18(-6) | 1.33(-4) | 6.26(-8) | 6.13(-6) |
| | *mean* | *5.26(-4)* | *6.72(-3)* | *1.52(-4)* | *3.01(-3)* | *1.09(-6)* | *1.30(-4)* | *3.66(-8)* | *6.03(-6)* |
| 80 | S | 1.19(-4) | 1.18(-3) | 2.33(-5) | 3.65(-4) | 2.84(-8) | 3.59(-6) | n.e. | 4.08(-8) |
| | L | 1.19(-4) | 1.19(-3) | 2.34(-5) | 3.67(-4) | 4.61(-8) | 3.67(-6) | n.e. | 3.43(-8) |
| | *mean* | *1.19(-4)* | *1.19(-3)* | *2.34(-5)* | *3.66(-4)* | *3.73(-8)* | *3.63(-6)* | *n.e.* | *3.76(-8)* |
| 100 | S | 4.63(-5) | 4.19(-4) | 7.70(-6) | 1.03(-4) | 3.94(-9) | 3.50(-7) | n.e. | 1.73(-9) |
| | L | 4.78(-5) | 4.19(-4) | 7.89(-6) | 1.02(-4) | 6.00(-9) | 3.02(-7) | n.e. | 4.37(-10) |
| | *mean* | *4.71(-5)* | *4.19(-4)* | *7.80(-6)* | *1.03(-4)* | *4.97(-9)* | *3.26(-7)* | *n.e.* | *1.08(-9)* |

760  n.e.  not evaluated

761

762

763 **Figure captions**

764 **Fig.1** Pressure dependence of the vesicularity of an ascending magma as described by our simulated

765 MORB melt at 1873K and for different initial $CO_2$ contents at depth (namely 0.06, 0.3, 0.6, 1.6, 4.6,

766 9.5, 21 and 29 wt% corresponding to $P_{CO_2}^{sat}=$ 1, 5, 10, 20, 30, 50, 80 and 100 kbar). The full curves

767 correspond to a super saturation ratio $\alpha=1$, and the dotted curves to $\alpha=2$ (see text). Symbols represent

768 the vesicularity of MORB samples collected on the Atlantic (squares), Indian (triangles) and Pacific

769 (crosses) ridges (data are from Chavrit (2010), Hekinian et al. (2000) and Pineau et al. (2004)). Values

770 of the initial $CO_2$ contents (in wt% $CO_2$) are indicated along the curves.

771 **Fig.2** Vesicularity of our simulated MORB evaluated at 1,000 m.b.s.l. (upper set of red dots, color

772 online) and at 5,000 m.b.s.l. (lower set of red dots, color online) as function of the initial $CO_2$ content

773 before degassing (the corresponding $CO_2$-saturation pressure is indicated on the upper axis). Data for

774 MORB samples dredged between ~1,000 m.b.s.l. and 5,400 m.b.s.l. on the Atlantic (squares), Indian

775 (triangles), and Pacific (crosses) ridges are shown for comparison (data source: Chavrit (2010) and

776 Pineau et al. (2004)). Notice that the $CO_2$ content in the (bulk) saturated melt at 1,000 m.b.s.l. is about

777 50 ppmw, and is about 250 ppmw at 5,000 m.b.s.l., values which become non negligible with respect

778 to the CO$_2$ content in vesicles only when the vesicularity is very low (V*$\leq 10^{-3}$).

779 **Fig.3** MORB-$CO_2$ noble gas partition coefficients as function of pressure. The dotted curves represent

780 the raw results obtained with small system sizes (lower curve) and with large system sizes (upper

781 curve), respectively, whereas the full curves are obtained from their arithmetic mean. The separation

782 between the two dotted curves gives an estimation of the error bar (see text and Table 2). Notice that

783 the distinction between dotted curves and full curves is barely visible for *He* and *Ne* over the entire

784 pressure range. For *Xe* the results become unreliable above 30 kbar and are not shown (see Table 2).

785 **Fig.4** $^4$*He* and $^{40}$*Ar* contents in melt (*m*) and in vesicles (*v*) as function of the vesicularity evaluated at

786 an hydrostatic pressure equivalent to 3,000 m.b.s.l.. The abundances in noble gases are normalized

787 with their initial values in the undegassed sample. The total $CO_2$ content of the MORB melt is
788 indicated by a number which corresponds to the $CO_2$-saturation pressure, $P_{CO_2}^{sat}$, given in kbar.

789 **Fig.5** $Ar$ content in vesicles (in $10^{-6}$ $g_{Ar}/g_{melt}$ or ppmw) as function of the vesicularity evaluated at
790 3,000 m.b.s.l.. The initial abundance in $^{40}Ar$ of the source region is assumed to be equal to that
791 measured in the *2πD43 popping rock* ( ~0.1ppmw, Moreira et al. (1998)) whereas different initial
792 conditions of $CO_2$-saturation are investigated (namely $P_{CO_2}^{sat}$= 0.4, 0.5, 0.6, 0.7, 0.8, 0.9, 1, 2, 3, 4, 5, 10,
793 20, 30, 40 and 50 kbar). The numbers associated with arrows indicate the value of $P_{CO_2}^{sat}$. The curve
794 labeled $\mathbf{V^1}$, and corresponding to the red dots (color on line), describes a closed system degassing
795 without vesicle loss. The curves labeled $\mathbf{V^2}$ (blue dots) describe a 2-stage vesiculation process with
796 vesicule loss. For illustration, three degassing paths are shown: ($\mathbf{P_1}$=20; $\mathbf{P_2}$=19, 18,.., 0.4 kbar), ($\mathbf{P_1}$=5;
797 $\mathbf{P_2}$=4,3,.., 0.4 kbar) and ($\mathbf{P_1}$=1; $\mathbf{P_2}$= 0.9, 0.8,.., 0.4 bar) where $\mathbf{P_1}$ is the pressure at which the first
798 vesiculation stage occurs and $\mathbf{P_2}$ the pressure of the second vesiculation stage, knowing that in
799 between the ascending magma has experienced a total loss of vesicles. For 3-stage vesiculation
800 (labeled $\mathbf{V^3}$ and green dots), and 4-stage vesiculation (labeled $\mathbf{V^4}$ and cyan dots), for the sake of clarity
801 only one degassing path is shown: ($\mathbf{P_1}$=20; $\mathbf{P_2}$=3; $\mathbf{P_3}$=2, 1,..,0.4 kbar) and ($\mathbf{P_1}$=20; $\mathbf{P_2}$=3; $\mathbf{P_3}$= 1: $\mathbf{P_4}$=0.9,
802 0.8,.., 0.4 kbar). Notice that most of the experimental data (black squares, data taken from Chavrit
803 (2010)) can be explained by a 2-, 3- or 4- stage vesiculation process, whereas the *2πD43 popping rock*
804 is well described by a closed system degassing starting at 10 kbar (i.e. ~30 km in the oceanic mantle).

805 **Fig.6** As in Fig.5 but for the $^4He/^{40}Ar$ ratio. The $^4He/^{40}Ar$ ratio of the source region is assumed to be
806 identical to that of the *2πD43 popping rock* (~1.5, see Moreira et al. (1998)).

807 **Fig.7** Snapshots of the simulation cell during a decompression experiment. Picture (a) shows a
808 snapshot of the simulation cell containing a MORB melt $CO_2$-saturated at 1873K and 50 kbar. The
809 system is composed of 5,000 silicate atoms, 275 $CO_2$ molecules (in fact ~99 $CO_2$ molecules and ~176
810 carbonate ions, $CO_3^{2-}$, on average) and one $He$ atom. For clarity, the size of the silicate atoms are
811 arbitrarily reduced with respect to that of the $CO_2$ molecules (carbon atoms are in blue, oxygen atoms

812    are in red and carbonate ions appear as bent $CO_2$ molecules) whereas the size of the *He* atom (in

813    yellow) is magnified. Notice that the distribution of $CO_2$ molecules within the melt is homogeneous

814    and that the *He* atom is not preferentially solvated by $CO_2$ molecules. The system is suddenly

815    decompressed from 50 to 5 kbar and it is observed, in <u>picture (b)</u>, at a time t=250 ps after the pressure

816    drop. On this snapshot most of the $CO_2$ molecules are forming clusters: one is visible near the center

817    of the box and another one, solvating the *He* atom, is located on the right side of the box. After 5,450

818    ps (<u>picture (c)</u>) all the nuclei have clustered together in a unique bubble in which the *He* atom is

819    enclosed. The final vesicularity of the system is about ~30% as expected from thermodynamics (see

820    Fig.1 for a 50→5 kbar decompression path).

821    **Fig.8** Kinetics of bubble formation and noble gas transfer during a decompression experiment. The

822    initial state is a well equilibrated MORB melt $CO_2$-saturated at 100 kbar and 1873K, and the final state

823    is a vesiculated melt at 10 kbar and 1873K. The decompression step (100→10 kbar) takes place at t=0

824    and is quasi instantaneous (the pressure drop is applied over one MD step, i.e. 0.001 ps). The figure

825    shows as function of running time the evolution of the atomic fraction in $CO_2$ molecules which are in

826    the immediate neighborhood of a $CO_2$ molecule, $f_{CC}$, of a *He* atom , $f_{HeC}$, and of an *Ar* atom, $f_{ArC}$. The

827    raw data are the wiggling curves (blue for $CO_2$, red for *He* and green for *Ar*, color online) whereas the

828    black dotted curves are least squared fits based upon a bi-exponential function ( see text and Eq.(10)).

829    The statistical noise of the raw data is much larger with *He* and *Ar* than with $CO_2$ because the number

830    of $CO_2$ molecules in the simulated sample is much larger than that of the noble gases (213 instead of

831    10). Notice that for t>0 the three functions $f_{CC}$, $f_{HeC}$, and $f_{ArC}$ exhibit a quasi identical steep rise at short

832    times without time lag, suggesting that bubble formation and noble gas catchment are simultaneous.

833    The time evolution of the vesicularity V* is also shown for comparison (thick black curve). Notice

834    that the vesicularity of the sample reaches a plateau value equal to 0.534 which is precisely the value

835    expected from thermodynamics at equilibrium (Eq.(7)). For a rapid check, read in Fig.1 the value of

836    the vesicularity at 10 kbar for an ascending melt saturated at 100 kbar (the curve marked 29 wt% in

837    Fig.1).

**Appendix**

838

839    To describe the noble gas-$CO_2$ interactions, we have used the molecular models for binary mixtures

840    developed by Vrabec et al. (2001, 2009) where the pair potential between a noble gas atom $X$ and a

841    $CO_2$ molecule is given by a sum of two Lennard-Jones (L-J) potentials,

842    $$u_{X,CO_2} = \sum_{n=1,2} 4\varepsilon_{XO} \left[\left(\frac{\sigma_{XO}}{r_{XO_n}}\right)^{12} - \left(\frac{\sigma_{XO}}{r_{XO_n}}\right)^6\right] \qquad (A1)$$

843    where $\varepsilon_{XO}$ and $\sigma_{XO}$ are L-J parameters and $r_{XO_n}$ is the distance between the noble gas atom $X$ and one

844    of the two oxygen atoms ($n$=1,2) belonging to the $CO_2$ molecule. The interaction between the noble

845    gas and the carbon atom is not considered because the latter one is deeply embedded into the

846    electronic clouds of the oxygen atoms. Trial values for $\varepsilon_{XO}$ and $\sigma_{XO}$ are deduced from the Lorentz-

847    Berthelot combining rules,

848    $$\sigma_{XO} = \frac{\sigma_X + \sigma_O}{2} \qquad (A2)$$

849    $$\varepsilon_{XO} = \sqrt{\varepsilon_X \varepsilon_O} \qquad (A3)$$

850    where $\sigma_X$, $\sigma_O$, $\varepsilon_X$ and $\varepsilon_O$ are L-J parameters describing the pure fluids (for *He*, see Dykstra (1989), and

851    for *Ne*, *Ar* and *Xe*, see Vrabec et al. (2001). Although the Lorentz-Berthelot combining rules are

852    widely used to model binary mixtures, there is room for improvement. So, we have followed the

853    procedure of Huang et al. (2009) which consists in introducing an adjustable parameter, $\xi_X$, in the

854    Lorentz-Berthelot rule for the pair energy parameter,

855    $$\varepsilon_{XO} = \xi_X \sqrt{\varepsilon_X \varepsilon_O}, \qquad (A4)$$

856    to reproduce at best Henry's constant of noble gases in liquid $CO_2$.

857    From a computational standpoint, Henry's constant, $H_X$, of a noble gas $X$ in a liquid solvent $S$ writes

858    (Shing et al., 1988)

$$H_X = \rho k_B T / \gamma_S^X = \rho k_B T \times e^{\frac{\mu_S^{X,ex}}{k_B T}} \qquad (A5)$$

where $\rho$ is the numerical density of the solvent (e.g. $CO_2$), T the temperature, $k_B$ the Boltzmann constant, and $\gamma_S^X$ the solubility parameter of the noble gas $X$ in the solvent. We have evaluated the Henry constant of He, Ne, Ar and Xe in liquid $CO_2$ at 273K by implementing the TPM (see Eq.(8)). Computational details are the same as those given in section 2.3. Thus, the simulated liquid sample is composed of 500 $CO_2$ molecules interacting through the Zhang and Duan potential (2005) modified by Guillot and Sator (2011) for bending, and the insertion energy of the noble gas atom in liquid $CO_2$ is calculated with the pair potential described above. For each noble gas, the Henry constant is evaluated via Eq.(A5) in adjusting by trial and error the parameter, $\xi_X$, to reproduce at best the noble gas solubility data of the literature. Final values of the L-J potential parameters for noble gas-$CO_2$ interactions are given in Table A1 as also as the comparison between calculated and measured Henry's constants.

**Table A1**

L-J potential parameters for noble gas-$CO_2$ interactions and comparison between calculated and experimental values for the Henry constant at 273K.

| $X$ | $\sigma_{XO}$(A) | $\varepsilon_{XO}$(kJ/mol) | $H_X^{calc}$(bar) | $H_X^{exp}$(bar) |
|-----|------|------|------|------|
| *He* | 2.81 | 0.525 | $2.0\pm0.2\ 10^3$ | [a]$2.05\pm0.35\ 10^3$ |
| *Ne* | 2.89 | 0.671 | $1.5\pm0.2\ 10^3$ | [b]$1.2\pm0.2\ 10^3$ |
| *Ar* | 3.19 | 1.131 | $3.0\pm0.3\ 10^2$ | [c]$3.4\pm0.6\ 10^2$ |
| *Xe* | 3.44 | 1.317 | $90\pm10$ | [d]87 |

a Mackendrick et al., 1968

b Sasinovskii, 1979

c Kaminishi et al., 1968

d Ackley and Notz, 1976

**BIBILIOGRAPHY**

Ackley R.D. and Notz K.J. (1976) Distribution of xenon between gaseous and liquid $CO_2$. Technical Report ORNL-5122, Oak Ridge National Lab. TN (USA), 1-35.

Albarède F. (2008) Rogue mantle helium and neon. Science **319**, 943-945.

Allègre C.J., Staudacher T., Sarda P. and Kurz M. (1983) Constraints on evolution of Earth's mantle from rare gas systematics. Nature **303**, 762-766.

Allègre C.J., Staudacher T. and Sarda P. (1986/87) Rare gas systematics, formation of the atmosphere, evolution and structure of the Earth's mantle. Earth Planet. Sci. Lett. **81**, 127-150.

Anderson D.L. (1998) The helium paradoxes. Proc. Natl. Acad. Sci. USA **95**, 4822-4827.

Aubaud C., Pineau F., Jambon A. and Javoy M. (2004) Kinetic disequilibrium of C, He, Ar and carbon isotopes during degassing of mid-ocean ridge basalts. Earth and Planet. Sci. Lett. **222**, 391-406.

Bauchy M., Guillot B., Micoulaut M. and Sator N. (2012) Viscosity and viscosity anomalies of model silicates and magmas: a numerical investigation. Chem. Geol., http://dx.doi.org/10.1016/J.chemgeo.2012.08.035..

Bologna M.S., Padilha A.L., Vitorello I. and Pádua M.B. (2011) Signatures of continental collisions and magmatic activity in central Brazil as indicated by a magnetotelluric profile across distinct tectonic provinces. Precambrian Res. **185**, 55-64.

Botcharnikov R., Freise M., Holtz F. and Behrens H. (2005) Solubility of C-O-H mixtures in natural melts: new experimental data and application range of recent models. Annals of Geophys. **48**, 633-646.

Bottinga Y. and Javoy M. (1990) MORB degassing: Bubble growth and ascent. Chem. Geol. **81**, 255-270.

Brooker R.A., Wartho J.-A., Carroll M.R., Kelley S.P. and Draper D.S. (1998) Preliminary UVLAMP determinations of argon partition coefficients for olivine and clinopyroxene grown from silicate melts. Chem. Geol. **147**, 185-200.

Brooker R.A., Kohn S.C., Holloway, J.R. and MCMillan P.F. (2001) Structural controls on the solubility of $CO_2$ in silicate melts Part I: bulk solubility data. Chem. Geol. **174**, 225-239.

Brooker R.A., Du Z., Blundy J.D., Kelley S.P., Allan N.L., Wood B.J., Chamorro E.M., Wartho J.-A. and Purton J.A. (2003) The "zero charge" partitioning behaviour of noble gases during mantle melting. Nature **423**, 738-741.

Burnard P. (1999) The bubble-by-bubble volatile evolution of two mid-ocean ridge basalts. Earth and Planet. Sci. Lett. **174**, 199-211.

Burnard P. (2001) Correction for volatile fractionation in ascending magmas; noble gas abundances in primary mantle melts, Geochim. Cosmochim. Acta **65**, 2605-2614.

Burnard P. (2004) Diffusive fractionation of noble gases and helium isotopes during mantle melting. Earth Planet. Sci. Lett. **220**, 287-295.

Burnard P., Graham D. and Farley K. (2004) Fractionation of noble gases (He, Ar) during MORB mantle melting: a case study on the Southeast Indian Ridge. Earth Planet. Sci. Lett. **227**, 457-472.

Cartigny P., Pineau F., Aubaud C. and Javoy M. (2008) Towards a consistent mantle carbon flux estimate: Insights from volatile systematics ($H_2O$/Ce, $\delta D$, $CO_2$/Nb) in the North Atlantic mantle (14°N and 34°N). Earth Planet. Sci. Lett. **265**, 672-685.

Chamorro E.M., Wartho J.-A., Brooker R.A., Wood B.J., Kelley S.P. and Blundy J.D. (2002) Ar and K partitioning between clinopyroxene and silicate melt to 8 GPa. Geochim. Cosmochim. Acta **66**, 507-519.

Chavrit D. (2010) Global mapping of *$CO_2$* fluxes along the Mid-Ocean Ridge system: a petrological approach. PhD thesis, Université de Nantes (France).

Clague D.A., Paduan J.B. and Davis A.S. (2009) Widespread strombolian eruptions of mid-ocean ridge basalt. J. Volcanol. Geotherm. Res. **180**, 171-188.

Colin A., Burnard P.G., Graham D.W. and Marrocchi Y. (2011) Plume-ridge interaction along the Galapagos Spreading center: discerning between gas loss and source effects using neon isotopic compositions and [4]He-[40]Ar*-$CO_2$ relative abundances. Geochim. Cosmochim. Acta **75**, 1145-1160.

Dalton J.A. and Presnall D.C. (1998) The continuum of primary carbonatitic-kimberlitic melt compositions in equilibrium with lherzolite: Data from the system CaO-MgO-$Al_2O_3$-$SiO_2$-$CO_2$ at 6 GPa. J. Petrol. **39**, 1953-1964.

Dasgupta R. and Hirschmann M.M. (2006) Melting in the Earth's deep upper mantle caused by carbon dioxide. Nature **440**, 659-662.

Deitrick G.L., Scriven L.E. and Davis H.T. (1989) Efficient molecular simulation of chemical-potentials. J. Chem. Phys. **90**, 2370-2385.

Dixon J.E., Stolper E. and Delaney J.R. (1988) Infrared spectroscopic measurements of $CO_2$ and $H_2O$ in Juan de Fuca Ridge basaltic glasses. Earth Planet. Sci. Lett. **90**, 87-104.

Dixon J.E., Stolper E.M. and Holloway J.R. (1995) An experimental study of water and carbon dioxide solubilities in mid-ocean ridge basaltic liquids Part I: Calibration and solubility models. J. Petrol. **36**, 1607-1631.

Dixon J.E. (1997) Degassing of alkalic basalts. Am. Mineral. **82**, 368-378.

Dixon J., Clague D.A., Cousens B., Monsalve M.L. and Uhl J. (2008) Carbonatite and silicate melt metasomatism of the mantle surrounding the Hawaiian plume: Evidence from volatiles, trace elements, and radiogenic isotopes in rejuvenated-stage lavas from Niihau, Hawaii. Geochem. Geophys. Geosyst. **9**, Q09005-1-34

Dykstra C.E. (1989) Molecular mechanics for weakly interacting assemblies of rare gas atoms and small molecules. J. Am. Chem. Soc. **111**, 6168-6174.

Eggler D.H. (1976) Does $CO_2$ cause partial melting in the low-velocity layer of the mantle ? Geology **4**, 69-72.

Eissen J.-Ph, Fouquet Y., Hardy D. and Ondréas H. (2003) Recent MORB volcaniclastic explosive deposits formed between 500 and 1750 m.b.s.l. on the axis of the mid-atlantic ridge, south of the Azores. In *Explosive Subaqueous Volcanism*, Geophys. Monograph **140**, 143-166.

Evans R.L., Hirth G., Baba K., Forsyth D., Chave A. and Mackie R. (2005) Geophysical evidence from the MELT area for compositional controls on oceanic plates. Nature **437**, 249-252.

Gaillard F., Malki M., Iacono-Marziano G., Pichavant M. and Scaillet B. (2008) Carbonatite melts and electrical conductivity in the astenosphere. Science **322**, 1363-1365.

Gardner J.E., Hilton M. and M.R. Carroll (2000) Bubble growth in highly viscous silicate melts during continuous decompression from high pressure. Geochim. Cosmochim. Acta **64**, 1473-1483.

Gonnermann H.M. and Mukhopadhyay S. (2009) Preserving noble gases in a convecting mantle. Nature **459**, 560-564.

Graham D. and Sarda P. (1991) Mid-ocean ridge popping rocks: implications for degassing at ridge crests - Comment. Earth Planet. Sci. Lett. **105**, 568-573.

970    Graham D.W. (2002) Noble gas isotope geochemistry of mid-ocean ridge and ocean island basalts:
971    Characterization of mantle source reservoirs. In *Noble Gases in Geochemistry and Cosmochemistry*
972    (eds. D. Porcelli, C.J. Ballentine and R. Wieler), Reviews in Mineralogy and Geochemistry **47**, 247-
973    319.

974    Guillot B. and Sarda P. (2006) The effect of compression on noble gas solubility in silicate melts and
975    consequences for degassing at mid-ocean ridges. Geochim. Cosmochim. Acta **70**, 1215-1230.

976    Guillot B. and Sator N. (2007a) A computer simulation study of natural silicate melts. Part I: Low
977    pressure properties. Geochim. Cosmochim. Acta **71**, 1249-1265.

978    Guillot B. and Sator N. (2007b) A computer simulation study of natural silicate melts. Part II: High
979    pressure properties. Geochim. Cosmochim. Acta **71**, 4538-4556.

980    Guillot B. and Sator N. (2011) Carbon dioxide in silicate melts: A molecular dynamics simulation
981    study. Geochim. Cosmochim. Acta **75**, 1829-1857.

982    Guillot B. and Sator N. (2012) Noble gases in high-pressure silicate liquids: A computer simulation
983    study. Geochim. Cosmochim. Acta **80**, 51-69.

984    Hammouda T. (2003) High-pressure melting of carbonated eclogite and experimental constraints on
985    carbon recycling and storage in the mantle. Earth Planet. Sci. Lett. **214**, 357-368.

986    Harper C.L. Jr. and Jacobsen S.B. (1996) Noble gases and Earth's accretion. Science **273**, 1814-1818.

987    Heber V.S., Brooker R.A., Kelley S.P. and Wood B. (2007) Crystal-melt partitioning of noble gases
988    (helium, neon, argon, krypton, and xenon) for olivine and clinopyroxene. Geochim. Cosmochim. Acta
989    **71**, 1041-1061.

990    Hékinian R., Chaigneau M. and Cheminée J.-L. (1973) Popping rocks and lava tubes from the Mid-
991    Atlantic rift valley at 36°N,. Nature **245**, 371-373.

992    Hékinian R., Pineau F., Shilobreeva S., Bideau D., Garcia E. and Javoy M (2000) Deep sea activity on
993    the Mid-Atlantic Ridge near 34°50'N: Magma composition, vesicularity and volatile content. J.
994    Volcanol. Geotherm. Res. **98**, 49-77.

995    Helo C., Longpré M.-A., Shimizu N., Clague D.A. and Stix J. (2011) Explosive eruptions at mid-
996    ocean ridges driven by $CO_2$-rich magmas. Nature Geoscience **4**, 260-263.

997    Huang Y.-L., Miroshnichenko S., Hasse H. and Vrabec J. (2009) Henry's law constant from molecular
998    simulation: A systematic study of 95 systems. Int. J. Thermophys. **30**, 1791-1810.

Jambon A., Weber H. and Braun O. (1986) Solubility of He, Ne, Ar, Kr and Xe in a basalt melt in the range 1250-1600°C. Geochemical implications. Geochim. Cosmochim. Acta **50**, 401-408.

Javoy M. and Pineau F. (1991) The volatiles record of a "popping" rock from the Mid-Atlantic Ridge at 14°N: chemical and isotopic composition of gas trapped in the vesicles. Earth Planet Sci. Lett. **107**, 598-611.

Jendrzejewski N., Trull T.W., Pineau F. and Javoy M. (1997) Carbon solubility in mid-ocean ridge basaltic melt at low pressures (250-1950 bar). Chem. Geol. **138**, 81-92.

Kaminishi G.I., Arai Y., Saito S. and Maeda S. (1968) Vapor-liquid equilibria for binary and ternary mixtures containing carbon dioxide. J. Chem. Eng. Jpn **1**, 109-116.

Mackendrick R.F., Heck C.K. and Barrick P.L. (1968) Liquid-vapor equilibria of the helium-carbon dioxide system. J. Chem. Eng. Data. **13**, 352-353.

Marty B. and Tolstikhin I.N. (1998) $CO_2$ fluxes from mid-ocean ridges, arcs and plumes. Chem. Geol. **145**, 233-248.

Marty B. and Zimmermann L. (1999) Volatiles (He, C, N, Ar) in mid-ocean ridge basalts: Assessment of shallow-level fractionation and characterization of source composition. Geochim. Cosmochim. Acta **63**, 3619-3633.

Meibom A., Anderson D.L., Sleep N.H., Frei R., Chamberlain C.P., Hren M.T. and Wooden J.L. (2003) Are high [3]He/[4]He ratios in oceanic basalts an indicator of deep-mantle plume components ? Earth Planet. Sci. Lett. **208**, 197-204.

Moreira M., Kunz J. and Allègre C. (1998) Rare gas systematics in popping rock: isotopic and elemental compositions in the upper mantle. Science **279**, 1178-1181.

Moreira M. and Sarda P. (2000) Noble gas constraints on degassing process. Earth and Planet. Sci. Lett. **176**, 375-386.

Namiki A. and Manga M. (2008) Transition between fragmentation and permeable outgassing of low viscosity magmas. J. Volcanol. Geotherm. Res. **169**, 48-60.

Navon O., Chekmir A. and Lyakhovsky V. (1998) Bubble growth in highly viscous melts: theory, experiments, and autoexplosivity of dome lavas. Earth Planet Sci. Lett. **160**, 763-776.

Pan V., Holloway J.R. and Hervig R.L. (1991) The pressure and temperature dependence of carbon dioxide solubility in tholeiitic basalt melts. Geochim. Cosmochim. Acta **55**, 1587-1595.

Paonita A. and Martelli M. (2006) Magma dynamics at mid-ocean ridges by noble gas kinetic fractionation: Assessment of magmatic ascent rates. Earth Planet. Sci. Lett. **241**, 138-158.

Paonita A. and Martelli M. (2007) A new view of the He-Ar-$CO_2$ degassing at mid-ocean ridges: Homogeneous composition of magmas from the upper mantle. Geochim. Cosmochim. Acta **71**, 1747-1763.

Parman S.W., Kurz M.D., Hart S.R. and Grove T.L. (2005) Helium solubility in olivine and implications for high $^3He/^4He$ in ocean island basalts. Nature **437**, 1140-1143.

Pepin R.O. (2006) Atmospheres on the terrestrial planets: Clues to origin and evolution. Earth Planet. Sci. Lett. **252**, 1-14.

Pineau F., Shilobreeva S., Hékinian, Bideau D. and Javoy M (2004) Deep sea activity on the Mid-Atlantic Ridge near 34°50'N: a stable isotope (C, H, O) study. Chem. Geol. **211**, 159-175.

Porcelli D. and Wasserburg G.J. (1995) Mass transfer of helium, neon, argon, and xenon through a steady-state upper mantle. Geochim. Cosmochim. Acta **59**, 4921-4937.

Prousevitch A.A., Sahagian D.L. and Anderson A.T. (1993) Dynamics of diffusive bubble growth in magmas: Isothermal case. J. Geophys. Res. **98**, 22,283-22,307.

Prousevitch A.A. and Sahagian D.L (1998) Dynamics and energetics of bubble growth in magmas: Analytical formulation and numerical modeling. J. Geophys. Res. **103**, 18,223-18,251.

Raquin A., Moreira M.A. and Guillon F. (2008) He, Ne, Ar systematics in single vesicles: Mantle isotopic ratios and origin of the air component in basaltic glasses. Earth Planset Sci. Lett. **274**, 142-150.

Rooney T.O., Herzberg C. and Bastow I.D. (2012) Elevated mantle temperature beneath East Africa. Geology **40**, 27-30.

Ryan M.P. and Sammis C.G. (1981) The glass transition in basalt. J. Geophys. Res. **86**, 9519-9535.

Saal A.E., Hauri E.H., Langmuir C.H. and Perfit M.R. (2002) Vapour undersaturation in primitive mid-ocean-ridge basalt and the volatile content of Earth's upper mantle. Nature **419**, 451-455.

Sarda P. and Guillot B. (2005) Breaking of Henry's law for noble gas and $CO_2$ solubility in silicate melt under pressure. Nature **436**, 95-98.

Sarda P. and Graham D. (1990) Mid-ocean ridge popping rocks: implications for degassing at ridge crests. Earth Planet. Sci. Lett. **97**, 268-289.

1057  Sarda P. and Moreira M. (2002) Vesiculation and vesicle loss in mid-ocean ridge basalt glasses: He,
1058  Ne, Ar elemental fractionation and pressure influence. Geochim. Cosmochim. Acta **66**, 1449-1458.

1059  Sasinovskii V (1979) Trudy Moskovskogo Energeticheskogo Instituta **364**, 13-18.

1060  Shaw A.M., Behn M.D., Humphris S. E., Sohn R. A. and Gregg P.M. (2010) Deep pooling of low
1061  degree melts and volatile fluxes at the 85°E segment of the Gakkel Ridge: Evidence from olivine-
1062  hosted melt inclusions and glasses. Earth Planet. Sci. Lett. **289**, 311-322.

1063  Shing K.S., Gubbins K.E. and Lucas K. (1988) Henry constants in non-ideal fluid mixtures: computer
1064  simulation and theory. Mol. Phys. **65**, 1235-1252.

1065  Smith W. and Forester T. (1996) DL_POLY_2.0: a general-purpose parallel molecular dynamics
1066  simulation package. J. Mol. Graph. **14**, 136-141.

1067  Sohn R.A., Willis C., Humphris S., Shank T.M., Singh H., Edmonds H.N., Kunz K., Hedman U.,
1068  Helmkr E., Jakuba M., Liljebladh B., Linder J., Murphy C., Nakamura K.-I, Sato T., Schlindwein V.,
1069  Stranne C., Tausenfreund M., Upchurch L., Winsor P., Jakobsson M and Soule A.(2008) Explosive
1070  volcanism on the ultralow-spreading Gakkel ridge, Arctic Ocean. Nature **453**, 1236-1238.

1071  Sparks R.S.J. (1978) The dynamics of bubble formation and growth in magmas: A review and
1072  analysis. J. Volcanol. Geotherm. Res. **3**, 1-37.

1073  Stolper E. and Holloway J.R. (1988) Experimental determination of the solubility of carbon dioxide in
1074  molten basalt at low pressure. Earth Planet. Sci. Lett. **87**, 397-408.

1075  Thomsen T.B. and Schmidt M.W. (2008) Melting of carbonated pelites at 2.5-5.0 GPa, silicate-
1076  carbonatite liquid immiscibility, and potassium-carbon metasomatism of the mantle. Earth Planet Sci.
1077  Lett. **267**, 17-31.

1078  Toramaru A. (1995) Numerical study of nucleation and growth of bubbles in viscous magmas. J.
1079  Geophys. Res. **100**, 1913-1931.

1080  Vrabec J., Stoll J. and Hasse H. (2001) A set of models for symmetric quadrupolar fluids. J. Phys.
1081  Chem. B **105**, 12126-12133.

1082  Vrabec J., Huang Y.-L. and Hasse H. (2009) Molecular models for 267 binary mixtures validated by
1083  vapor-liquid equilibria: A systematic approach. Fluid Phase Equilib. **279**, 120-135.

1084  Walker D. and Mullins O. (1981) Surface tension of natural silicate melts from 1,200°-1,500°C and
1085  implications for melt structure. Contrib. Mineral. Petrol. **76**, 455-462.

1086    Widom B. (1963) Some topics in theory of fluids. J. Chem. Phys. **39**, 2808-2812.

1087    Yamamoto J., Nishimura K., Sugimoto T., Takemura K., Takahata N. and Sano Y. (2009) Diffusive
1088    fractionation of noble gases in mantle with magma channels: Origin of low He/Ar in mantle-derived
1089    rocks. Earth Planet. Sci. Lett. **280**, 167-174.

1090    Zeng G., Chen L.-H., Xu X.-S., Jiang S.-Y. and Hofmann A.W. (2010) Carbonated mantle sources for
1091    Cenozoic intra-plate alkaline basalts in Shandong, North China. Chem. Geol. **273**, 35-45.

1092    Zhang Z. and Duan Z. (2005) An optimized molecular potential for carbon dioxide. J. Chem. Phys.
1093    **122**, 214507-1-15.

1094

1095

1096

1097

1098

1099

1100

1101

1102

1103

1104

1105

1106

1107

1108

1109

1110

1111

1112

1113

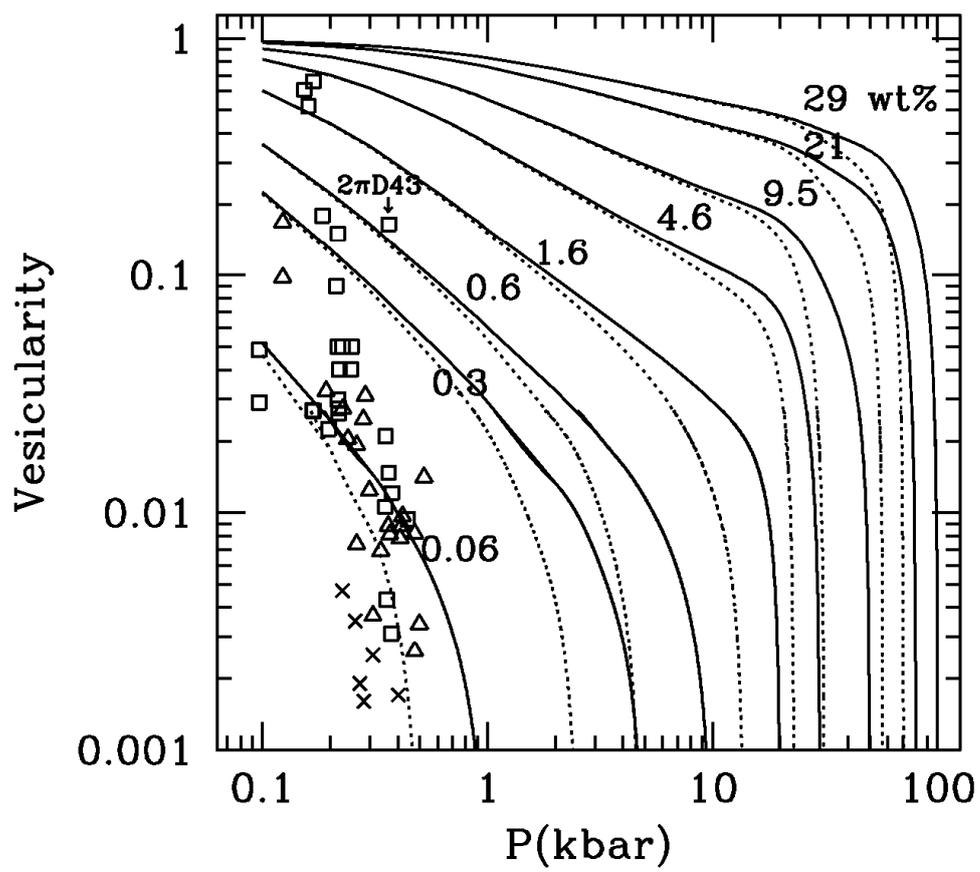

Fig.1

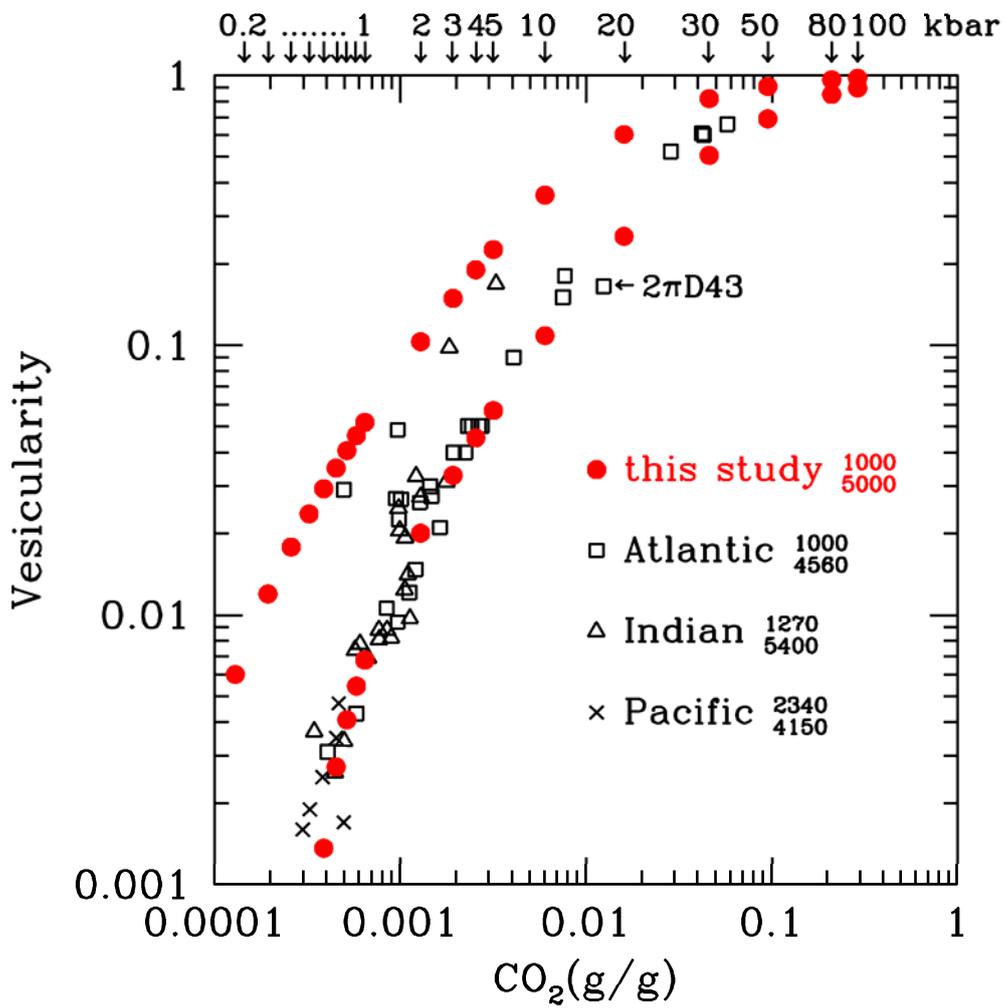

Fig.2

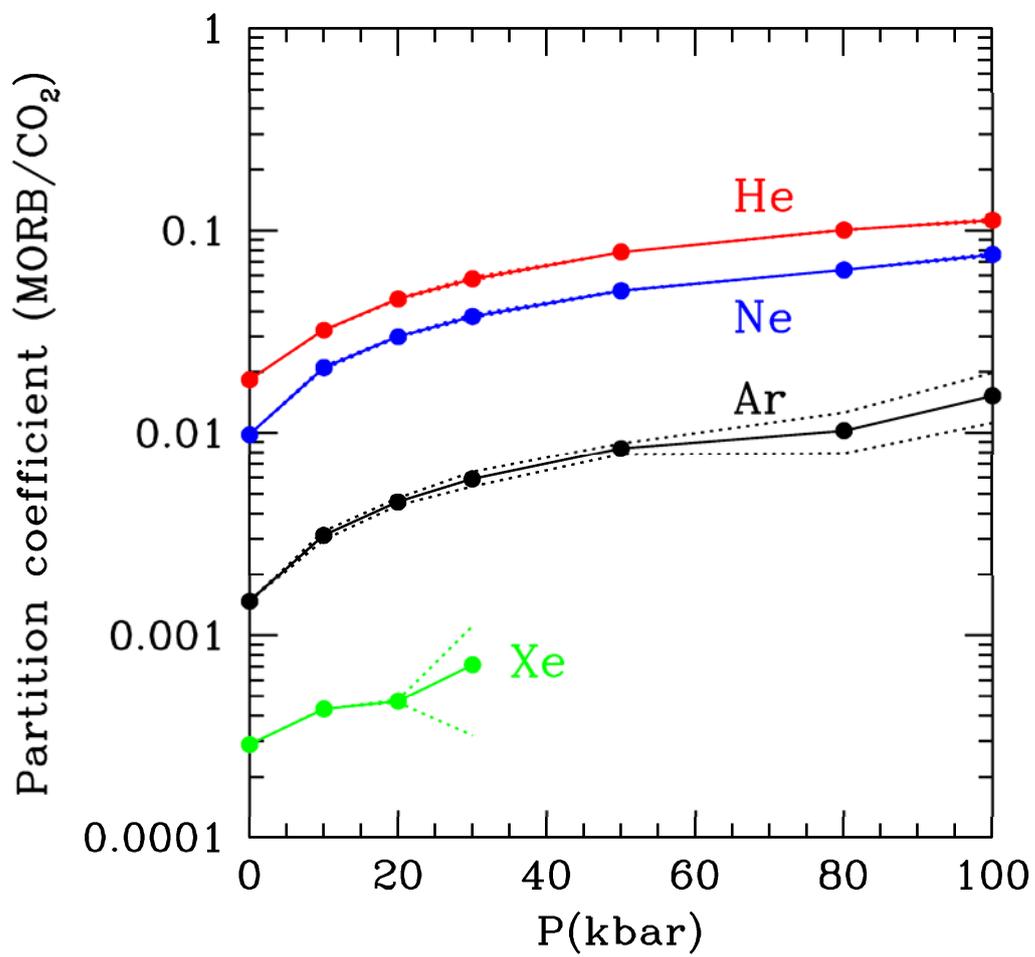

Fig.3

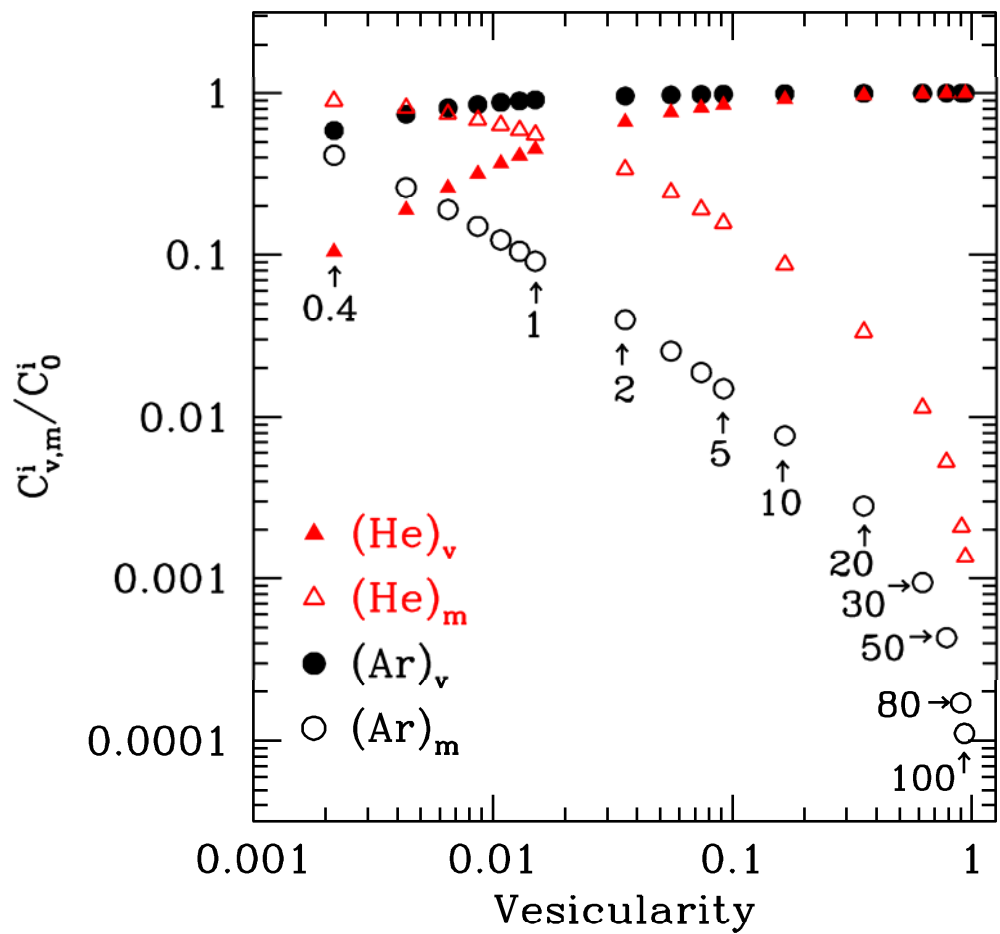

Fig.4

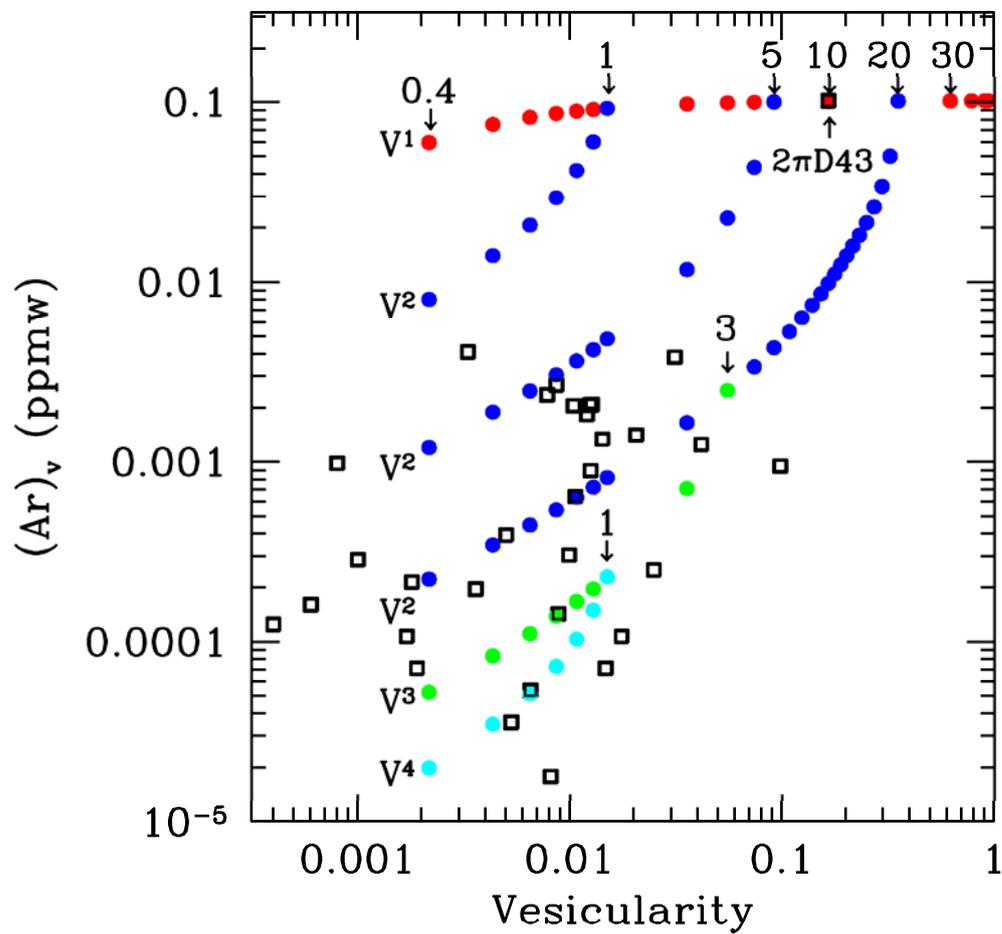

1143

Fig.5

1144

1145

1146

1147

1148

1149

1150

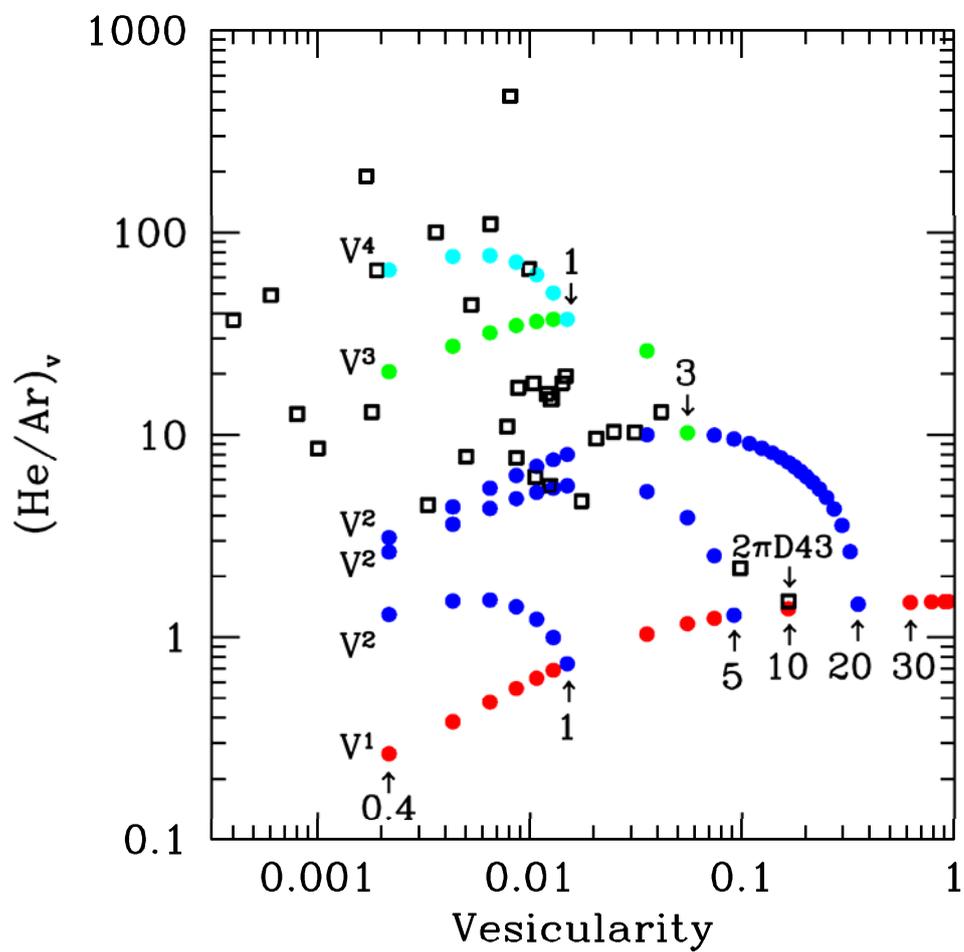

Fig.6

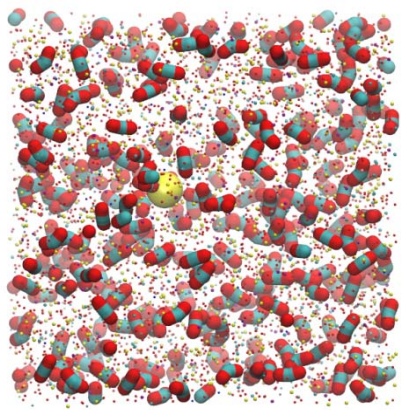

1157                    (a)

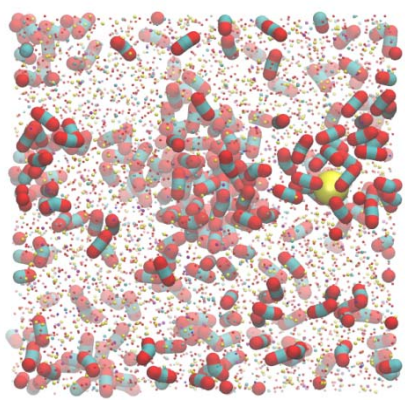

1158                    (b)

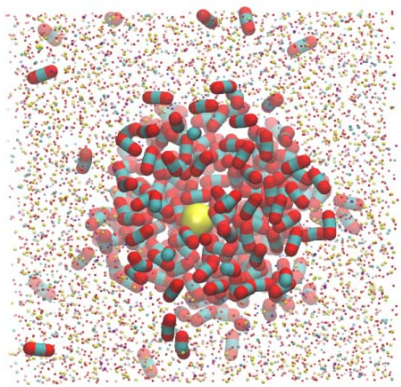

1159                    (c)                    Fig.7

1160

1161

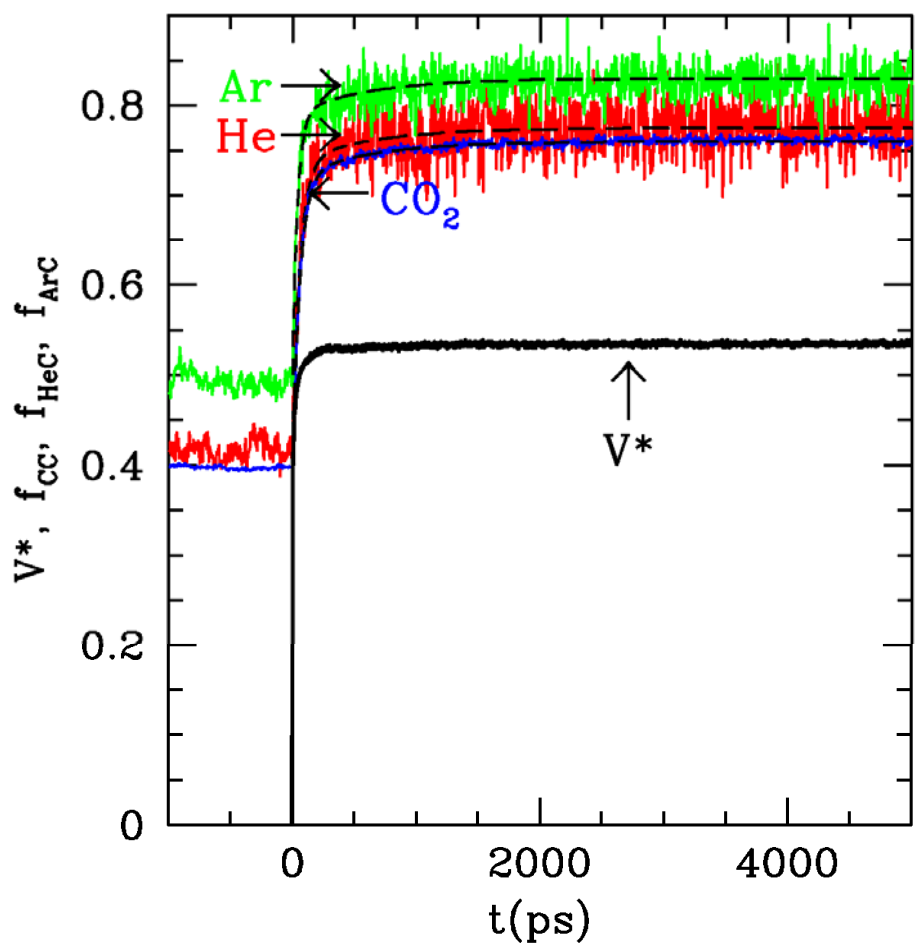

Fig.8

1162

1163

1164

1165

1166